\documentclass[aps, prl,
 amsmath,
 amssymb,
reprint,
showkeys,
]{revtex4-1}
\usepackage{graphicx}
\usepackage{epstopdf}
\usepackage{subfig}
\usepackage[geometry]{ifsym}
\usepackage{dcolumn}
\usepackage{bm}
\usepackage{caption}
\begin{document}
\title{Evolution and stability of shock waves in dissipative gases characterized by activated inelastic collisions}
\author{N. Sirmas}
   \email{nsirmas@uottawa.ca}
\author{M. I. Radulescu}
   \email{matei@uottawa.ca}
\affiliation{Department of Mechanical Engineering, University of Ottawa}
\date{\today}

\begin{abstract}
Previous experiments have revealed that shock waves driven through dissipative gases may become unstable, for example, in granular gases, and in molecular gases undergoing strong relaxation effects. The mechanisms controlling these instabilities are not well understood. We successfully isolated and investigated this instability in the canonical problem of piston driven shock waves propagating into a medium characterized by inelastic collision processes.  We treat the standard model of granular gases, where particle collisions are taken as inelastic with constant coefficient of restitution.  The inelasticity is activated for sufficiently strong collisions.  Molecular dynamic simulations were performed for 30,000 particles.  We find that all shock waves investigated become unstable, with density non-uniformities forming in the relaxation region.  The wavelength of these fingers is found comparable to the characteristic relaxation thickness. Shock Hugoniot curves for both elastic and inelastic collisions were obtained analytically and numerically. Analysis of these curves indicate that the instability is not of the Bethe-Zeldovich-Thompson or Dyakov-Kontorovich types. Analysis of the shock relaxation rates and rates for clustering in a convected fluid element with the same thermodynamic history outruled the clustering instability of a homogeneous granular gas. Instead, wave reconstruction of the early transient evolution indicates that the onset of instability occurs during the re-pressurization of the gas following the initial relaxation of the medium behind the lead shock. This re-pressurization gives rise to internal pressure waves in the presence of strong density gradients.  This indicates that the mechanism of instability is more likely of the vorticity-generating Richtmyer-Meshkov type, relying on the action of the inner pressure waves development during the transient relaxation.     

\end{abstract}

\maketitle

\section{I.~Introduction}

\begin{figure}[bp]
\centering
\includegraphics[width=.95\linewidth]{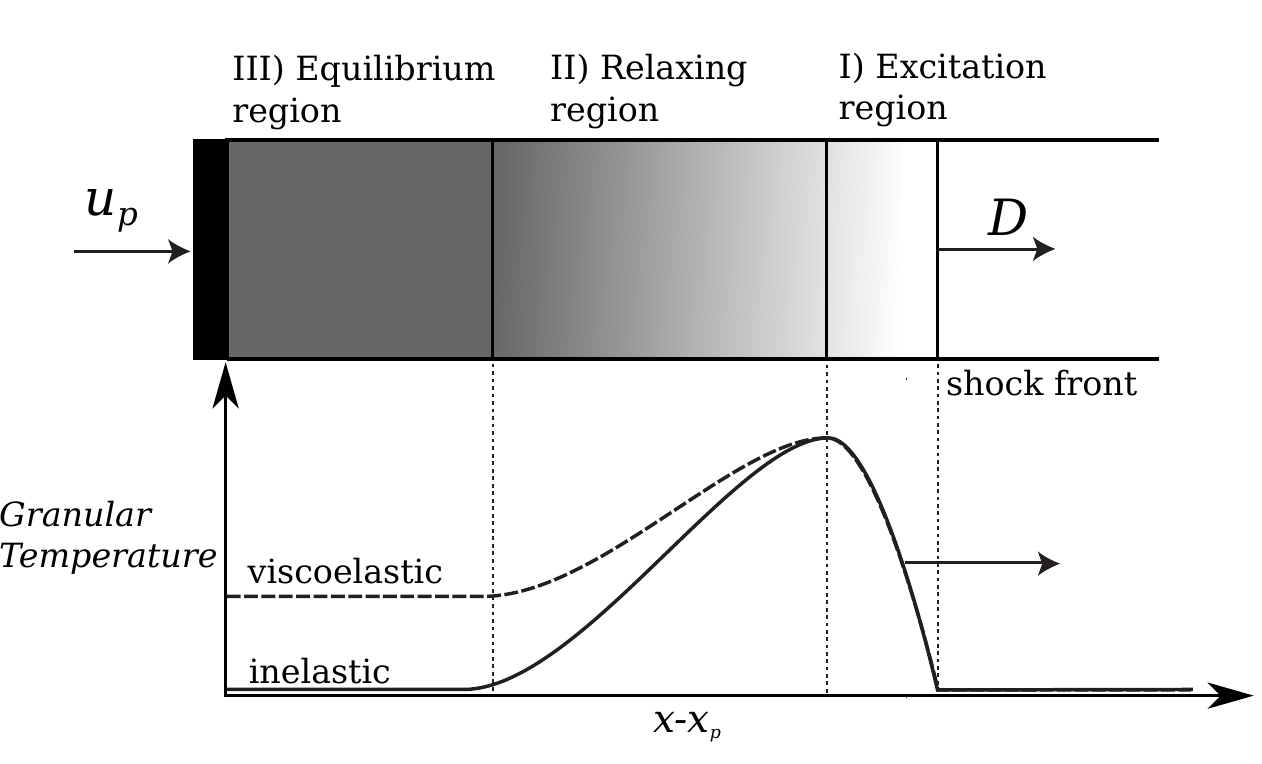}
\caption{Temperature distribution for a thermally relaxing shock wave travelling at velocity $D$}\label{fig:TemperatureProfile}
\end{figure}

Shock waves driven into dissipative gases sometimes develop instabilities.  Granular media, which are characterized by inelastic particle collisions, is one example. Previous experiments have identified unstable formations of finger-like jets in granular media dispersed by shock waves driven through air~\cite{Frost2012, Rodriguez2013} and for rapid granular flows down a chute~\cite{Boudet2013}. Similar pattern formations can be seen when granular media are subjected to a vertically oscillating bed, both experimentally and numerically~\cite{Bizon1998, Carrilo2008}. In the latter, the periodic agitation of the container walls drive strong shocks and expansion waves into the non-uniform granular gas.  The complex transient dynamics involved in the these past configurations have prevented the authors to clearly identify the mechanisms controlling the instability.  In the present study, we pose the problem in the classical formulation of a piston suddenly accelerated to a constant speed into a gas medium, as illustrated in Figure~\ref{fig:TemperatureProfile}.

Previous investigations of this canonical problem have looked at the one-dimensional structure and evolution of shock waves in granular gases, although instabilities had not been identified \cite{Goldshteinetalch31996, Kamenetsky_etal2000}. Goldshtein \textit{et al.} revealed that the structure of shock waves driven by a piston into a granular gas is composed of three distinct regions~\cite{Goldshteinetalch31996}. The first region follows the shock front, and is composed of a rapid increase in granular temperature (region I). Due to the inelasticity and increased rate of the collisions within this excited region, the granular temperature of the material falling further behind the shock starts to decrease, while density increases; this marks the `relaxing' region (region II). Eventually, the collision amplitudes become sufficiently weak such that visco-elastic particles collide elastically.  In this `equilibrium' region (region III), the gas retains a finite granular temperature. When all collisions are assumed inelastic, the equilibrium region tends to zero granular temperature.  

Kamenetsky \textit{et al.}~\cite{Kamenetsky_etal2000} investigated the evolution of such a structure numerically by solving the one-dimensional Euler equations for granular media. The authors revealed interesting dynamics prior to the shock wave attaining the developed structure illustrated in Figure~\ref{fig:TemperatureProfile}. In particular, the authors found that the lead shock front pulls back towards the piston for a short period, before attaining a constant velocity. The dynamics of this stage were not explained nor further explored.  Nevertheless, as we will show in the present article, these turn out to have a strong bearing on the multi-dimensional shock instability.



Qualitatively, a structure similar to that shown in Figure~\ref{fig:TemperatureProfile} is observed for sufficiently strong shock waves driven into molecular gases, whereby the shock is strong enough to bring about inelastic collisions between molecules (i.e., via endothermic reactions)~\cite{Zeldovich&Raizer1966}. Interestingly, these types of relaxing shock waves have also been shown to sometimes become unstable. Unstable shock structures have been observed experimentally in sufficiently strong shocks leading to ionization~\cite{Grunetal1991, Glass&Liu1978, Griffithetal1976}, dissociation~\cite{Griffithetal1976} and in gases with high specific heats~\cite{Griffithetal1976, Mishinetal1981,Hornung&Lemieux2001, Semenov2012}.

Current models for predicting such shock instability are mostly based on jump conditions between the initial and final equilibrium states, without knowledge of the kinetic processes linking the two states.  The D'Yakov-Kontorovich (DK) and the Bethe-Zel’dovich-Thompson (BZT) mechanisms require the shape of the Hugoniot curves to have anomalous properties (see, for example, Refs. \cite{Zeldovich&Raizer1966} and \cite{Landau&Lifshitz1987}). The Hugoniot curve is the locus of the equilibrium post shock state, usually represented in the pressure-specific volume plane.  While the Hugoniot curves can be obtained experimentally for a given substance, investigation of their properties in the context of BZT and DK instabilities predicted stable shocks at experimental conditions corresponding to unstable shocks~\cite{Hornung&Lemieux2001, Griffithetal1976}. 

Another mechanism of interest involved in shock instability is that of Richtmyer-Meshkov and Rayleigh-Taylor type instabilities, although such instabilities have not been reported in the cases above. In such a multi-dimensional instability, misaligned gradients of density and pressure lead to vorticity production~\cite{Brouillette2002}. This type of instability is a universal physical phenomena encountered, for example, in gases~\cite{Dimonte2010}, plasmas~\cite{Keskinen2006}, Bose-Einstein condensates~\cite{Bezett2010}, and combustion~\cite{Petchenkoetal2006}. 

Models for predicting the instability of relaxing shocks involving the kinetics of the relaxation process have only very recently been formulated.  Direct numerical simulations at the continuum level in the case of ionizing shocks has indeed recovered the instability \cite{Kapper&Cambier2011, Mondetal1997}, suggesting that it is related to the hydrodynamic coupling with the kinetics of the relaxation process.  This suggests that an account for the kinetics of the relaxation process may be required to predict the shock instability in relaxing media.



In the absence of bulk flow, it has been shown that such dissipative gases are subject to clustering instabilities \cite{Goldhirsch&Zanetti1993, Mitrano2011, Poscheletal2005}. This \textit{clustering instability}, first shown by Goldhirsch and Zanetti \cite{Goldhirsch&Zanetti1993}, is seen in granular gases, where the collisions can be assumed to remain inelastic for all impact conditions. In such a medium, an initially homogeneous gas develops clusters during its cooling, which takes the form of filamentous structures.  Gas is preferentially accelerated towards regions of higher density, owing to the local greater rate of pressure decay in these regions due to dissipation. Since the material passing through the shock structure undergoes the same cooling process, the clustering instability may be controlling the local non-homogeneities within the shock structure. This link is further explored in the present paper.    

To summarize, the goal of our present study is twofold.  We wish to first isolate the shock instability in relaxing media in a canonical problem, conducive to further analysis.  Second, we wish to determine the mechanism controlling the instability.  The qualitative correspondence of the structure of granular gases and molecular gases suggests that both problems can be studied by the same formalism, provided the collision properties are modified to account for the finite temperature equilibrium region of molecular gases. 

To investigate the evolution and stability of such shock waves, we adopt the simple kinetic model previously used to describe dissipative granular gases by Goldhirsch and Zanetti: the collision between ``hard particles'' of finite radius is modeled deterministically using a constant coefficient of restitution taken below unity. This model is the simplest kinetic model that can mimic relaxation. In order to capture the structure of relaxing gases more closely, we also assume that the collisions are activated by an impact energy threshold. Such a threshold is also applicable to granular media, which has been used to better imitate the visco-elastic behavior of colliding particles~\cite{Poscheletal2003}.

The paper is organized as follows: Section II outlines the details of the molecular dynamics model used in this study. Section III addresses the evolution and structure of shock waves predicted by the molecular dynamic model. Section IV provides further discussion and analysis of the mechanism controlling the shock instability. Finally, Section V offers our closing remarks.

\section{II.~Details of Molecular Dynamics Model}\label{sec:MDdetails}

The approach we use is a deterministic hard particle dynamic approach in a 2D environment, akin to the probabilistic approach of the Direct Simulation Monte Carlo (DSMC) technique \citep{Bird1999}.  In such models, only the collision rules are prescribed in order to capture a physical phenomenon (granular gases, relaxation, chemical reactions, etc.).  We employ the standard deterministic method used for granular gases, both in its kinetic theory and in particle based simulations. Indeed, much of the kinetic theory of dilute, idealized gases can be obtained by treating molecules as hard spheres with no internal structure \cite{Chapman&Cowling1970, Vincenti&Kruger1975}.

The current model assumes that collisions with boundaries are elastic, yielding a symmetry condition that is implemented in order to not artificially introduce supplementary system size effects. Each binary collision is elastic, unless an \textit{activation} threshold is reached. The post-collision velocities of two particles are calculated as:  
\begin{equation}\label{eq:postcoll}
\begin{aligned}
{\vec{u}_{i}} '=&\vec{u}_{i}-\frac{1}{2}(1+\varepsilon^*)\vec{g}_{ij}^n\\
\vec{u}_{j}'=&\vec{u}_{j}+\frac{1}{2}(1+\varepsilon^*)\vec{g}_{ij}^n\\
\end{aligned}
\end{equation}
where $\vec{g}_{ij}^n=\vec{u}_{i}^n-\vec{u}_{j}^n$ is the normal component of the relative velocity of the two disks.

Activation is assumed to occur when the collision between two disks is sufficiently strong.  This mimics the excitation of higher degrees of freedom (rotation, vibration, dissociation, ionization, etc.) with increasing temperatures \cite{Vincenti&Kruger1975}. This is also a simple model for granular media undergoing visco-elastic collisions~\cite{Poscheletal2003}.  Quantitatively, the collision between two disks is assumed to be elastic if $\vec{g}_{ij}^n$ is below a threshold ${u^*}$, a classical activation formalism in chemical kinetics. For collisions with a higher amplitude, we assume an inelastic dissipative collision, which is modeled with a constant coefficient of restitution $\varepsilon<1$. i.e.,
\begin{equation}
  \varepsilon^* = \left\{
  \begin{array}{l l}
    1 & \quad \text{if\quad  $|\vec{g}_{ij}^n|<{u^*}$}\\
    \varepsilon & \quad \text{if\quad  $|\vec{g}_{ij}^n|\geq {u^*}$}\\
  \end{array} \right.
\end{equation}
where the predefined ${u^*}$ and $\varepsilon$ remain constant during each simulation.

The problem we study is a classical shock propagation problem, whereby the motion of a suddenly accelerated piston driven in a thermalized medium drives a strong shock wave. The driving piston is initially at rest and suddenly acquires a constant velocity $u_p$. Collisions with the piston are elastic.  This model allows for the dissipation of the non-equilibrium energy accumulated within the shock structure, which terminates once the collision amplitudes fall back below the activation threshold.  In this manner, the activation threshold also acts as a tunable parameter to control the equilibrium temperature in the post shock media.  Note that the model assumed is also the standard model for granular gases \cite{Brilliantov&Poschel2004}, allowing us to compare with the established hydrodynamic description of this type of media.

The MD simulations thus reconstruct the dynamics of hard disks.  These are calculated using the Event Driven Molecular Dynamics (EDMD) technique first introduced by Alder and Wainright \cite{Alder&Wainright1959}.  We use the implementation of P\"{o}schel and Schwager \cite{Poschel&Schwager2005}, that we have extended to treat a moving wall (piston). The particles were initialized with equal speed and random directions.  The system was let to thermalize and attain Maxwell-Boltzmann statistics.  Once thermalized, the piston started moving with constant speed. This code was implemented and tested for non-dissipative media in our previous study \cite{Sirmasetal2012}, where the simulated shock jump conditions agreed with those which were derived for hard disk mixtures.  

The initial packing factor of the disks was chosen to be $\eta_1=(N\pi d^2)/4A=0.012$, where $N\pi d^2/4$ is the volume (area) of the $N$ hard disk with diameter $d$, and $A=L_x\times L_y$ the domain area; the initial gas is thus in the ideal gas regime \cite{Sirmasetal2012}. 
All distances have been normalized by the initial mean free path of the system of disks $\lambda_1$, which takes the form \cite{Brilliantov&Poschel2004}: 
\begin{equation}
\lambda_1=\frac{1}{\sqrt{2\pi}dn_1b_2(\eta_1)}
\end{equation}
where $b_2(\eta)=(1-\frac{7\eta}{16})/(1-\eta)^2$ is the Enskog factor for a 2D system of hard particles, and $n_1=N/A$ is the initial number density of particles.
All speeds are scaled by the initial root mean squared velocity $u_{rms(1)}$ of the disks, fixing the time scaling by the \textit{initial} mean free time $\tau_1={\lambda_1}/{u_{rms(1)}}$. 

The numerical experiments were performed using 30,000 disks, unless otherwise noted.  A domain size of $L_x\times L_y=172.9\times17.2 $ and disk radius $\sigma=0.019$ was used to satisfy the packing factor of $\eta_1=0.012$.  The dimensions of the domain, with 30,000 particles, was found to be an appropriate size to investigate and capture instability, allowing for sufficiently fast computing in order for results to be ensemble averaged. Ensemble and coarse grain averaging was implemented to investigate the one-dimensional shock structure. For each set of parameters, an ensemble of 50 simulations was taken, with the macroscopic properties taken in strips of width $\Delta x\approx 0.5 \lambda_1$ parallel to the piston face. 

All macroscopic properties are scaled by the initial state, unless otherwise noted. 
The density $\rho$ is taken by tracking the number of disks within each strip, and the granular temperature is taken with the root mean squared velocity, i.e., $T=\frac{1}{2}u_{rms}^2$. The pressure is approximated from the Helfand equation of state for elastic disks~\cite{Sirmasetal2012}:
\begin{equation}
p=\frac{\rho T}{(1-\eta)^2}
\end{equation}


%
To investigate the dynamics of the shock waves, the family of characteristics were constructed. The particles paths ($P$), forward ($C^+$) and backward ($C^-$) running characteristics on an $x$ vs. $t$ plane are given by:
\begin{equation}\label{eq:char}
P:~\frac{dx_p}{dt}=u~~~~~C^+:~\frac{dx_+}{dt}=u+c~~~~~C^-:~\frac{dx_-}{dt}=u-c
\end{equation}
where $u$ is the local particle velocity normal to the piston and $c$ is the local speed of sound, at a given time. They represent the trajectories of fluid particles, right running pressure waves and left running pressure waves, respectively~\cite{Landau&Lifshitz1987}. The scaled speed of sound for such a media is approximated for an elastic system of disks, taken as~\cite{Sirmasetal2012}:
\begin{equation}
\frac{c}{u_{rms_1}}=\sqrt{\frac{T}{T_1}(1+(1-\eta)^{-2}+2\eta(1-\eta)^{-1})}
\end{equation}
The local packing factor is taken from the density jump, $\eta=\eta_1 {\rho}/{\rho_1}$.

The trajectories of the characteristics were obtained numerically by integrating \eqref{eq:char}. The $C^+$ characteristics are initiated from the piston face at specified intervals in time, while $C^-$ characteristics are initiated from the shock front at similar time intervals.  Particle paths are initialized at specified locations away from the initial piston position, denoted as $\xi=x(t=0)$ for each path.

\section{III.~Results}\label{sec:structure}

In this section we discuss the results obtained using the described model. We compare the evolution of shock structure and ensuing instability for varying properties. First, we look at the evolution of shock structure in detail for a single case. Next, we perform a parametric study to see how the evolution, shock structure, and stability vary with $u_p$, $u^*$ and $\varepsilon$.  

\subsection{A.~Evolution of shock structure}
 \begin{figure}[bp]
 \centering
 \includegraphics[width=\linewidth]{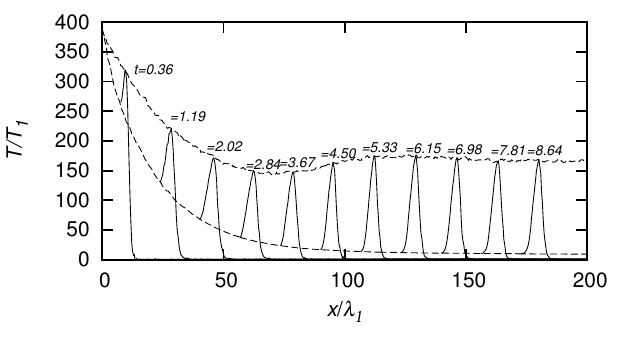}
 \caption{Evolution of shock structure for $u_p=20$, $u^*=10$, and $\varepsilon=0.95$.}
 \label{fig:eps95_u20_evolution_with_labels}
 \end{figure}

   \begin{figure*}[htbp]
   	\captionsetup[subfigure]{ oneside,margin={-1cm,2cm}, type=figure} 
   	\subfloat[]{\includegraphics[trim=0.1cm .5cm 0cm 0cm, clip=true, width=0.33\textwidth]{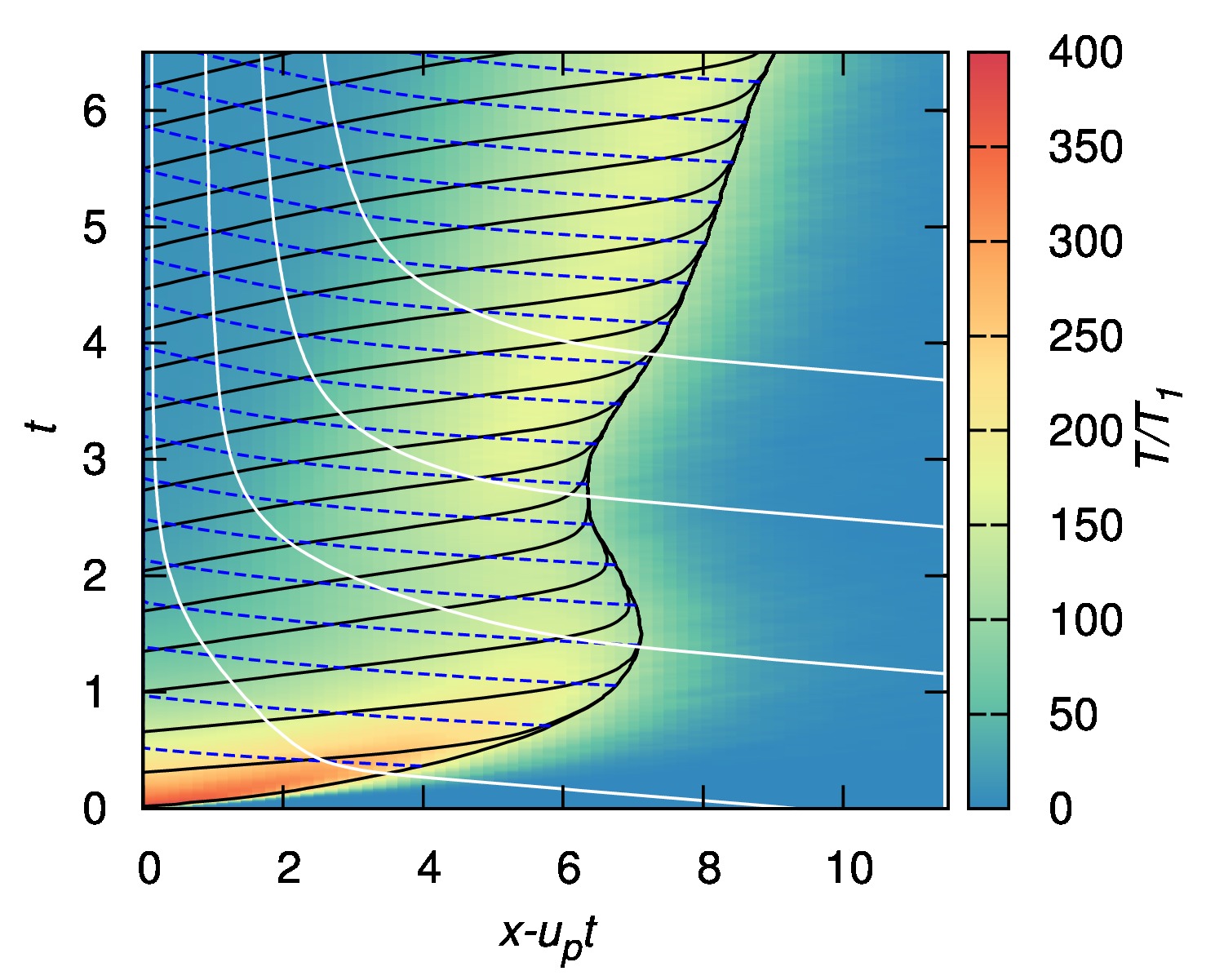}}
   	\subfloat[]{\includegraphics[trim=0.1cm .5cm 0cm 0cm, clip=true, width=0.33\textwidth]{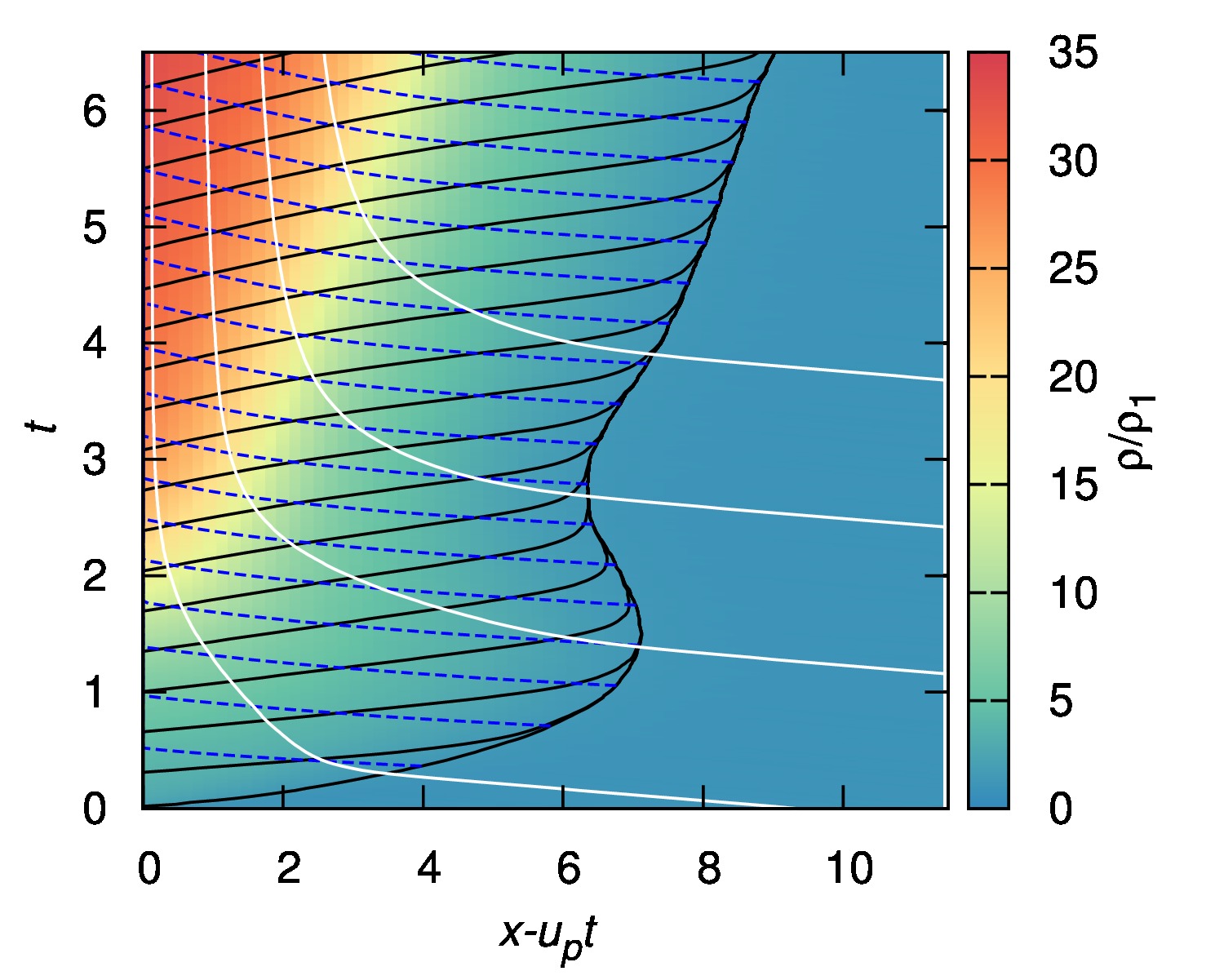}}
   	\subfloat[]{\includegraphics[trim=0.1cm .5cm 0cm 0cm, clip=true, width=0.33\textwidth]{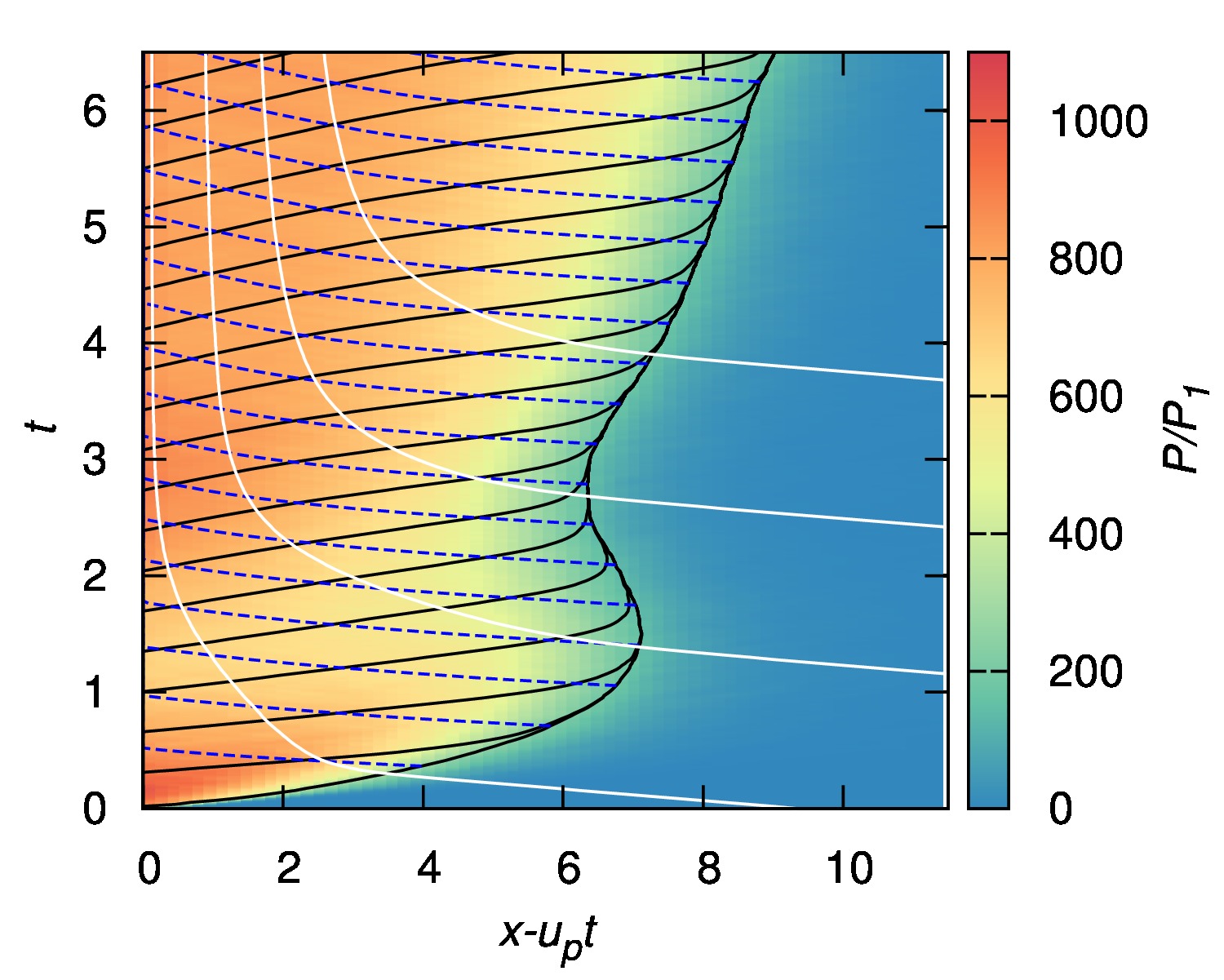}}
   	\caption{(Colour online) Evolution of (a) temperature, (b) density and (c) pressure on an $x$ vs.~$t$ plane, in the piston frame of reference, for $u_p=20$, $u^*=10$ and $\varepsilon=0.95$. Evolutions shown with select particle paths (solid white), forward (solid black) and backward (dashed blue) running characteristics.}
   	\label{fig:xt_pframe_MD}
   \end{figure*}
  \begin{figure*}[htbp]
  	\captionsetup[subfigure]{ oneside,margin={0cm,0cm}, type=figure} 
  	\subfloat[]{\includegraphics[trim=.5cm .5cm 0cm 1cm, clip=true, width=0.23\textwidth]{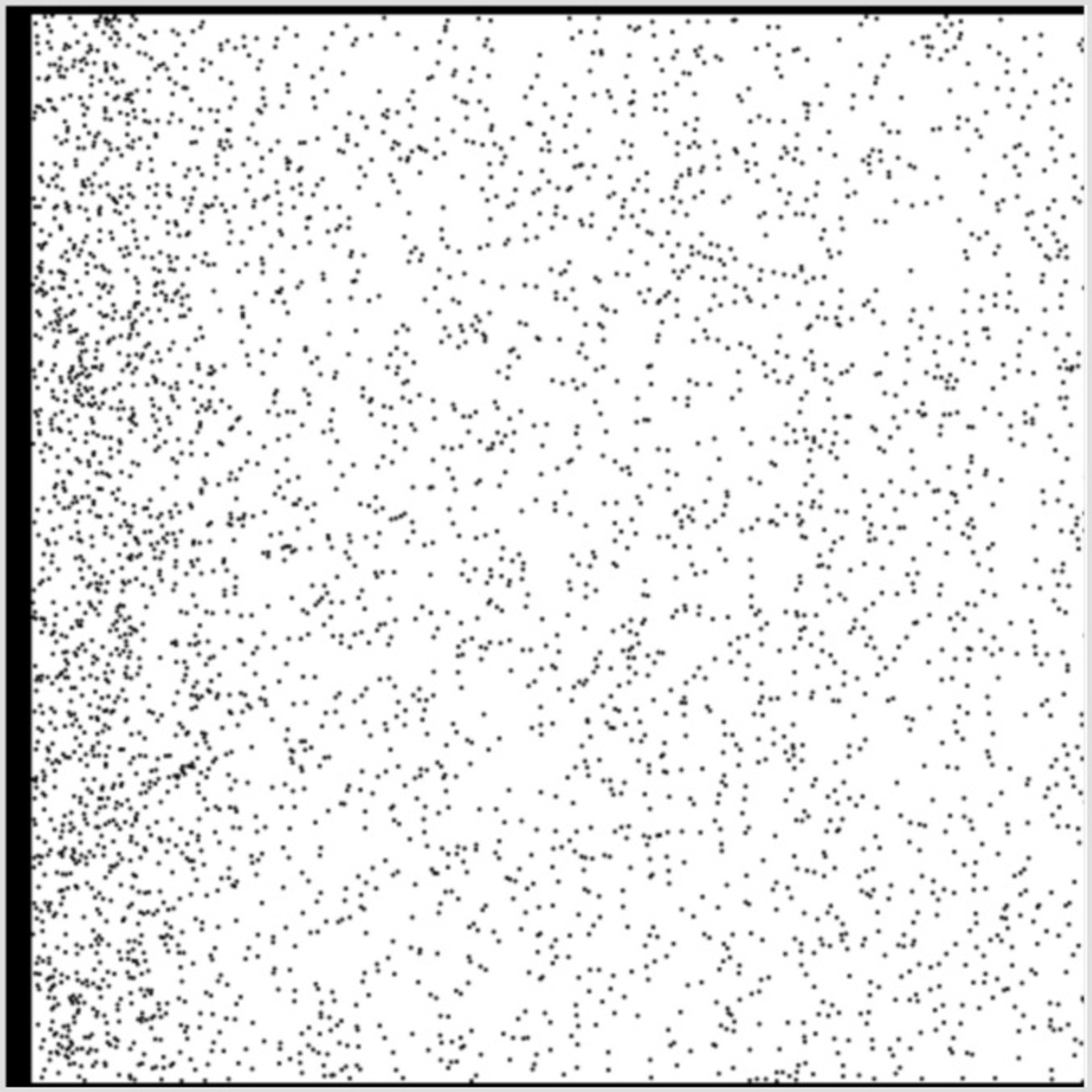}}~
  	\subfloat[]{\includegraphics[trim=.5cm .5cm 0cm 1cm, clip=true, width=0.23\textwidth]{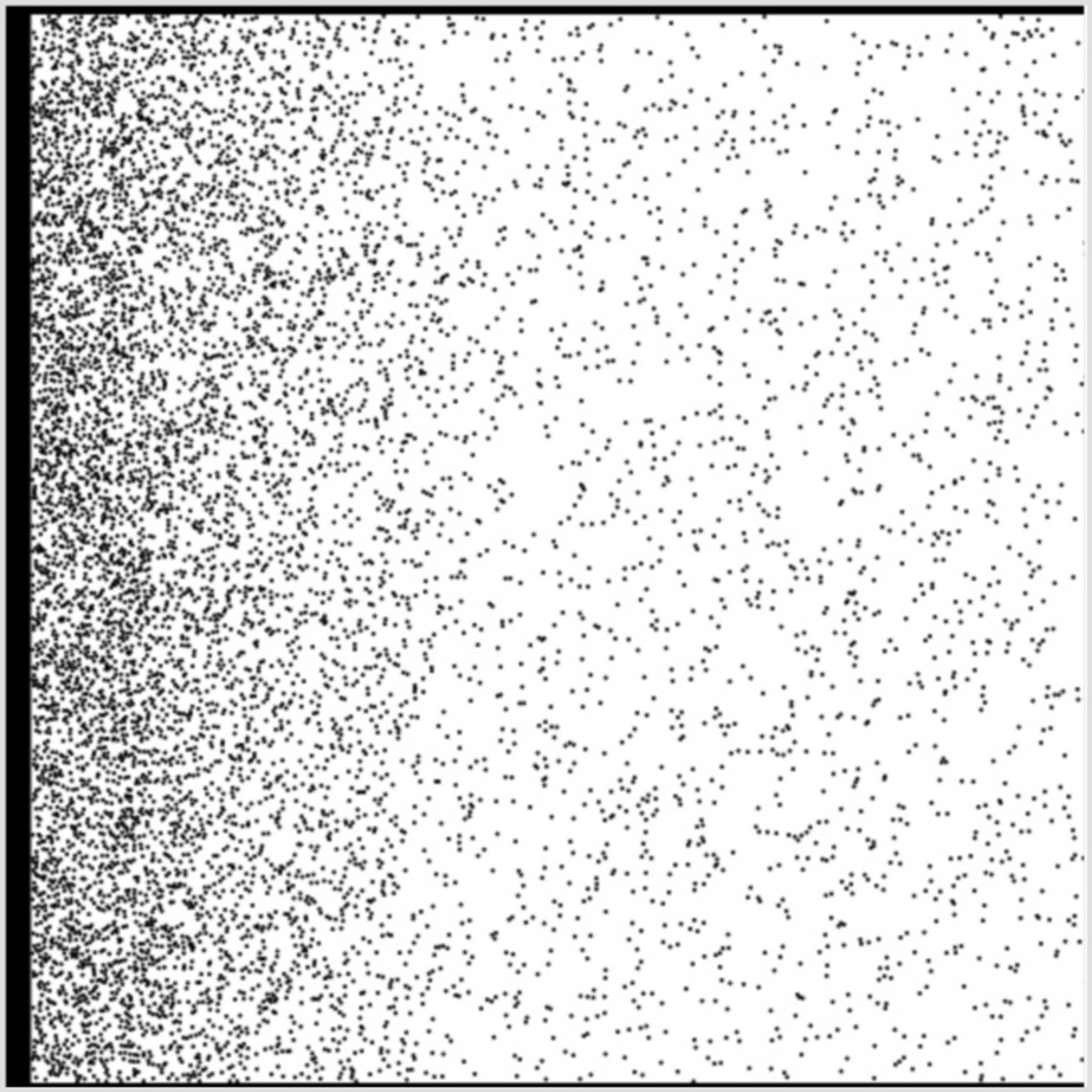}}~
  	\subfloat[]{\includegraphics[trim=.5cm .5cm 0cm 1cm, clip=true, width=0.23\textwidth]{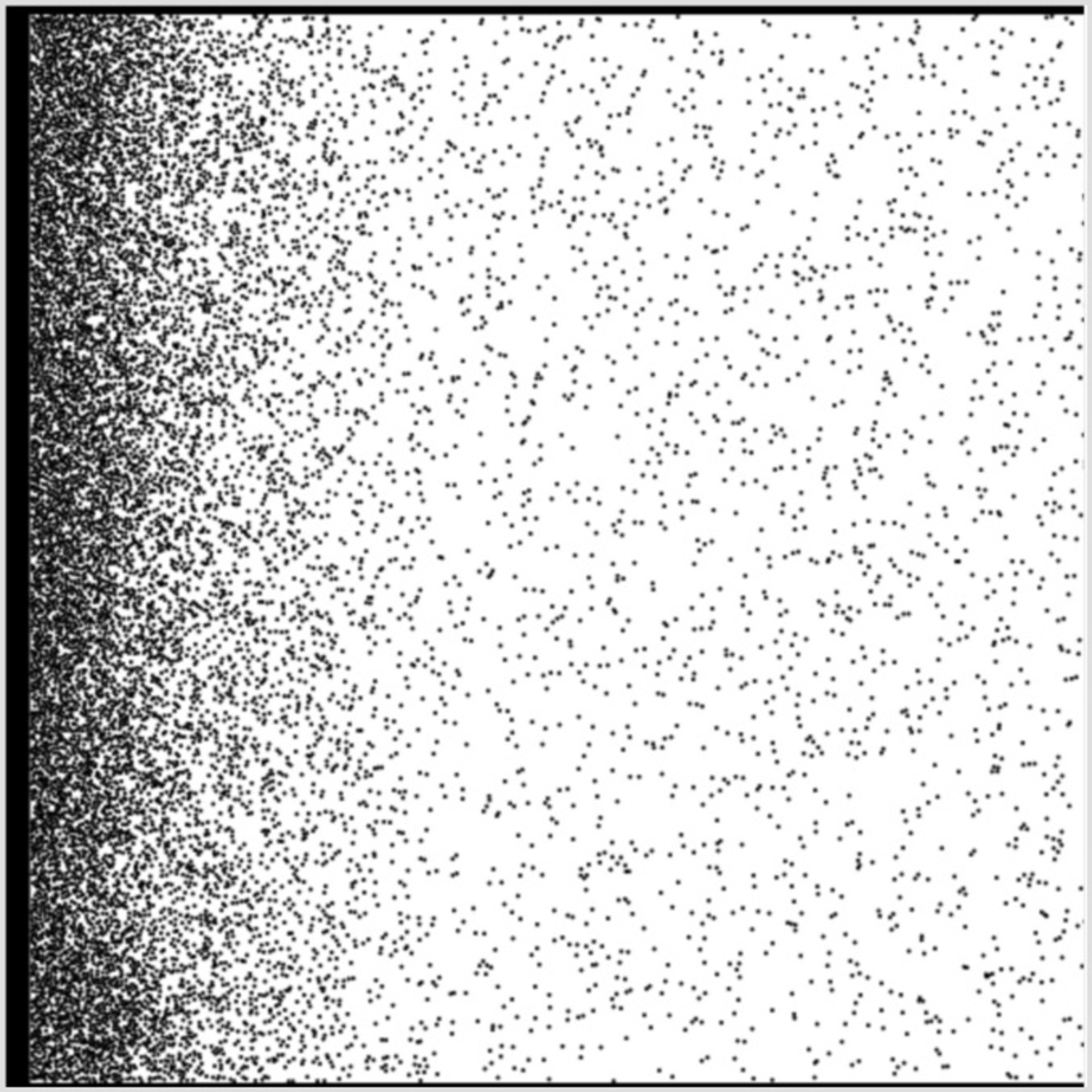}}~
  	\subfloat[]{\includegraphics[trim=.5cm .5cm 0cm 1cm, clip=true, width=0.23\textwidth]{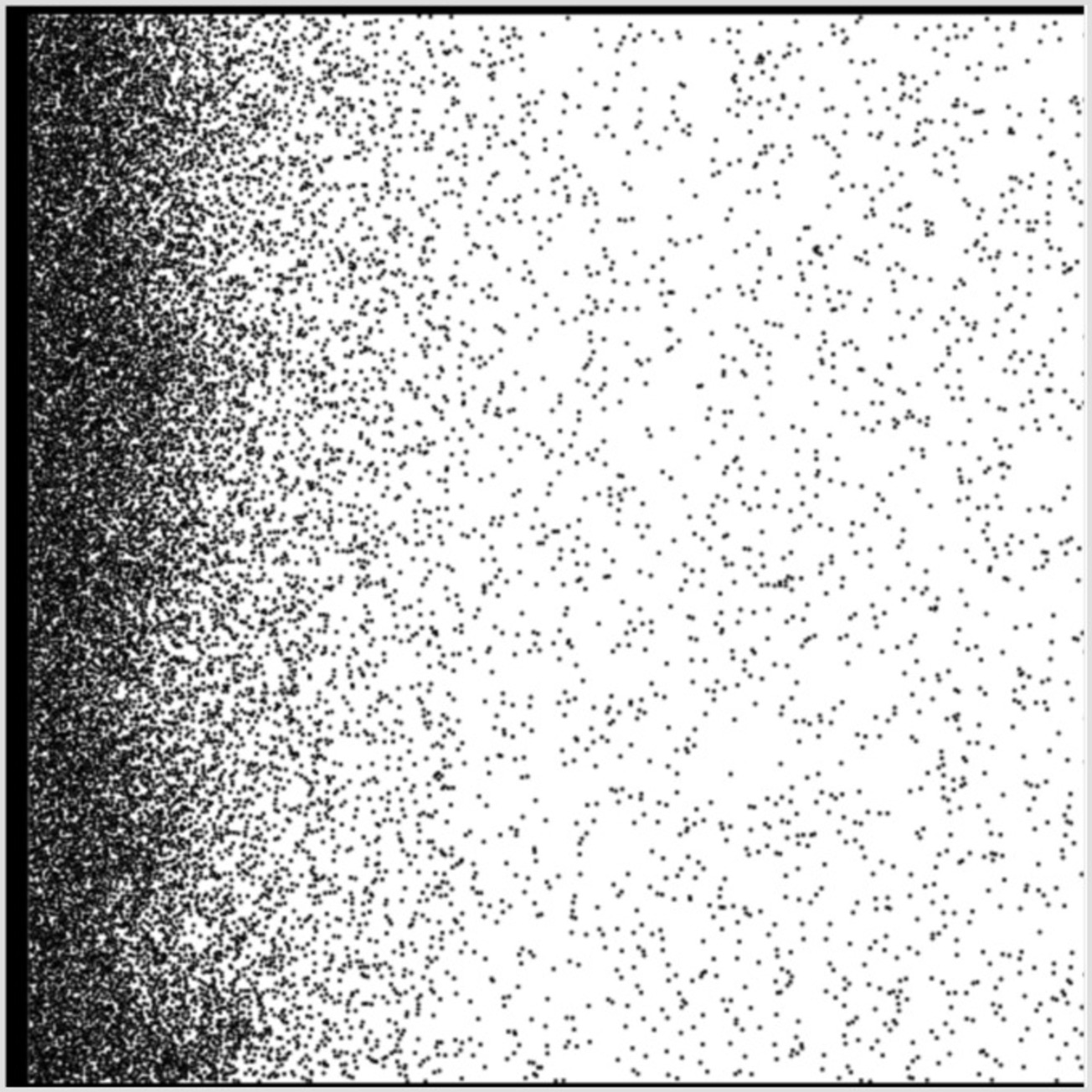}}~
  	\caption{Evolution of shock morphology, for a single realization, at  $t=$ (a) 0.3, (b) 1.5, (c) 2.7 and (d) 3.9; with $u_p=20$, $u^*=10$ and $\varepsilon=0.95$.}
  	\label{fig:evolution2D}
  \end{figure*}
  
The first case we look at is for $u_p=20$, $u^*=10$, and $\varepsilon=0.95$. Figure~\ref{fig:eps95_u20_evolution_with_labels} shows an example of the evolution of the one-dimensional temperature distribution. In addition to showing the instantaneous structure, the peak temperature and temperature at the piston are tracked. Initially the temperature jump of the shock is approximately $u_p^2\approx400$, as predicted for a system of elastic disks~\cite{Sirmasetal2012}. The temperature measured at the piston surface decays until coming to a quasi-equilibrium state, at which point most inelastic collisions have subsided - note that the kinetic model taken maintains an exponentially small fraction of activated collisions as the temperature decays below the activation temperature.  The peak temperature also decays initially, which is followed by an oscillation before reaching an equilibrium peak temperature. These dynamics are very similar to the ones predicted by Kamenetsky et al. in inviscid hydrodynamic simulations of granular gases with a constant $\varepsilon$ \cite{Kamenetsky_etal2000}. 

Figure~\ref{fig:xt_pframe_MD} shows the evolution of the averaged temperature, density and pressure fields in the $x$ - $t$ plane, in a frame of reference moving with the piston.  Selected particle paths, $C^+$ characteristics extending from the piston, and $C^-$ from the shock front are also shown in order to more clearly illustrate the dynamics. For example, the shock is the locus along which all forward facing pressure waves $C^+$ coalesce.  The shock wave driven by the piston generates an increase in the medium pressure, density and temperature.  As the medium behind the shock begins to cool, the lead shock is seen to decay. The cooling of the gas and decay of the lead shock can be correlated by the forward facing pressure waves.  The excess relaxation behind the lead shock leads to an eventual pull-back of the lead shock towards the piston.  A similar pullback was observed by Kamenetsky et al.~in their hydrodynamic simulations \cite{Kamenetsky_etal2000}. 

The cooling of the gas behind the lead shock, which can be followed along the corresponding particle paths, eventually is punctuated by an increase of density and a re-pressurization.  This can be clearly observed at $t \approx 2$.  The origin of this re-pressurization is not clear at present, but may be correlated with the arrival of the rear facing pressure waves (along the $C^-$ characteristics shown), originating at the decaying shock.  Interestingly, the rear re-pressurization leads to a forward-facing pressure wave, arriving at the lead shock at $t \approx 3$.  This marks the re-acceleration of the lead shock towards its final equilibrium structure.     

Figure~\ref{fig:evolution2D} shows the evolution of shock morphology for this case, obtained from a single realization. These results show the birth of an unstable structure, which we distinguish by density perturbations and corrugations appearing within the shock structure. Initial stages of the evolution do not show distinguishable instabilities, as seen at $t=0.3$, and up to $t=1.5$. This is the point where the shock front stops propagating ahead of the piston. For later times, instabilities in the form of high density clusters and corrugations appear at the piston face. This is seen at $t=2.7$, confirming that these instabilities occur between $t=1.5$ and $t=2.7$.

Comparing with the evolution of pressure shown in Figure \ref{fig:xt_pframe_MD}(c), this range in time is when the early particle paths undergo a re-pressurization event on route to attaining an equilibrium state. This indicates that the origin of the instability may be associated with this distinct feature of the relaxation process; a possible mechanism is discussed in Section IV. Once the shock evolution enters the developed stage, the clusters begin growing from the piston, as demonstrated by the snapshot at $t=3.9$.

\begin{figure}[tbp]
\includegraphics[width=.9\linewidth]{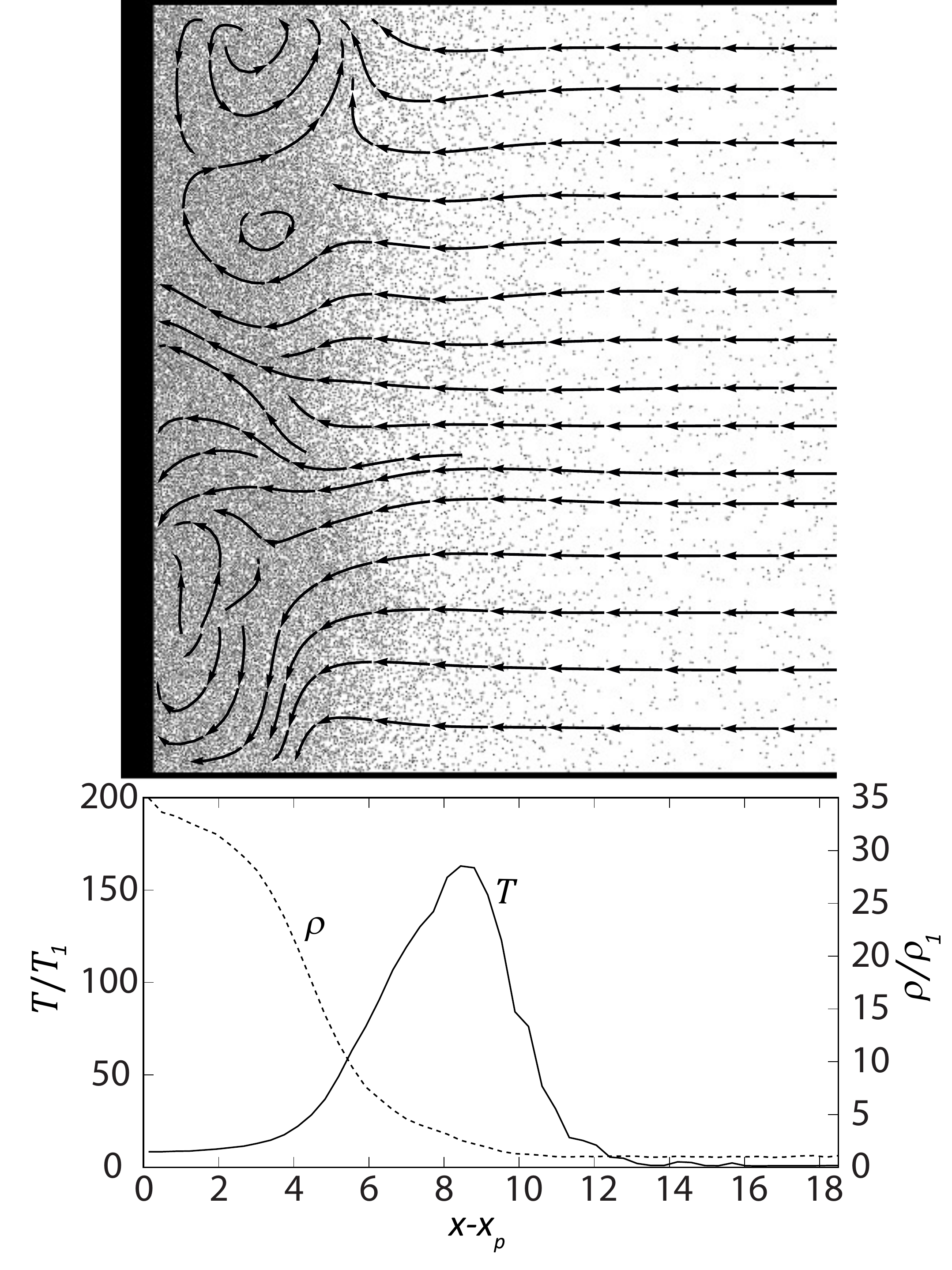}
\caption{Particle distribution and coarse-grained stream-lines for a single realization (top), with ensemble and coarse grained distributions of temperature and density (bottom) after $t=8.13$, with  $u_p=20$, ${u^*}=10$, and $\varepsilon=0.95$.}
\label{fig:Stream+profile}
\end{figure}
 
Figure~\ref{fig:Stream+profile} shows the particle distribution in the shocked material in relation to the mean temperature and density distributions.  Superposed on the particle distribution plot is the coarse grained velocity vector field.  This instantaneous vector field is rendered using streamlines, in order to better visualize the existence of coherent structures.  The streamlines were obtained by interpolating on the uniform grid of coarse grained averaged velocity vector field.  Results show that substantial disturbances in speed are present in the region of the high density gradients. Streamlines converge toward the high density fingers, giving rise to convective rolls.  


\subsection{B.~Parametric study of the shock structure and its evolution}
\subsubsection*{Dimensional analysis and independent parameters}
The macroscopic dynamics of the model introduced is expected to have a relatively small number of controlling parameters.  Dimensional analysis permits us to determine the number of parameters controlling the dynamics.  The initial thermodynamic state is uniquely defined by its granular temperature $T_1$, density $\rho_1$ and packing fraction $\eta_1$.  The shock dynamics, depend on the piston speed $u_p$, the activation threshold $u^*$ and the degree of inelasticity, $\varepsilon$.  Furthermore, we are interested at conditions in which the strong shock limit applies and the initial internal energy does not control the dynamics \cite{Zeldovich&Raizer1966, Sirmasetal2012}; this is the case where the experimental observations of shock instability have been made, for both the granular and relaxing molecular gases, as discussed in the Introduction. Under the scaling of our variables, this reduces to the limit where $u_p \gg 1$ and $u^* \gg 1$.  Under this limit, the parameters of the problem reduce to $u_p /u^*$, $\eta$, and $\varepsilon$. 

\begin{figure*}[htbp]
\captionsetup[subfigure]{oneside,margin={0.75cm,2.5cm}} 
\subfloat[]{\label{fig:equalratiod}\includegraphics[width=.45\linewidth]{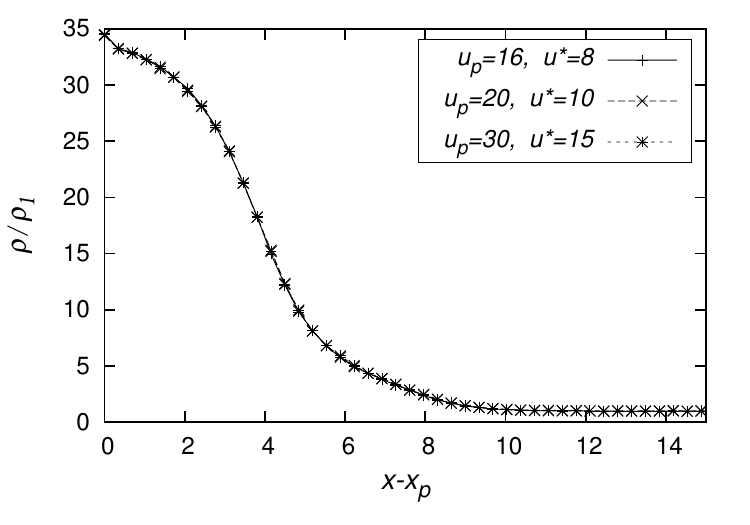}}\hfill
\subfloat[]{\label{fig:equalratioT}\includegraphics[width=.45\linewidth]{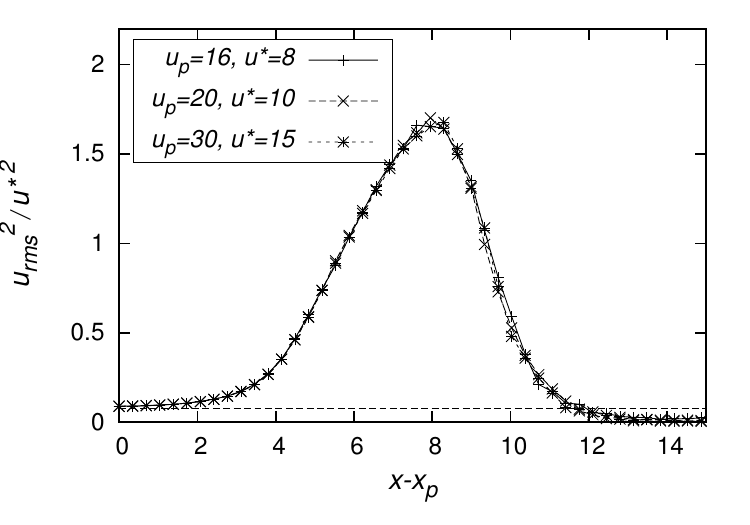}}
\caption{Ensemble and coarse grained one-dimensional shock structure (a) of density and (b) kinetic energy after a piston displacement of $x_p=138.7 \lambda_1$ for varying $u_p$ and $u^*$ with $u_p/{u^*}$=2, and $\varepsilon=0.95$.}
\label{fig:eq_ratio}
\end{figure*}
\begin{figure*}[htbp]
\captionsetup{type=figure}
\captionsetup[subfigure]{labelformat=empty, oneside,margin={-1cm,0cm}, type=figure} 
	\subfloat[]{\includegraphics[trim=0.1cm 0cm 0cm 0cm, clip=true, width=0.33\textwidth]{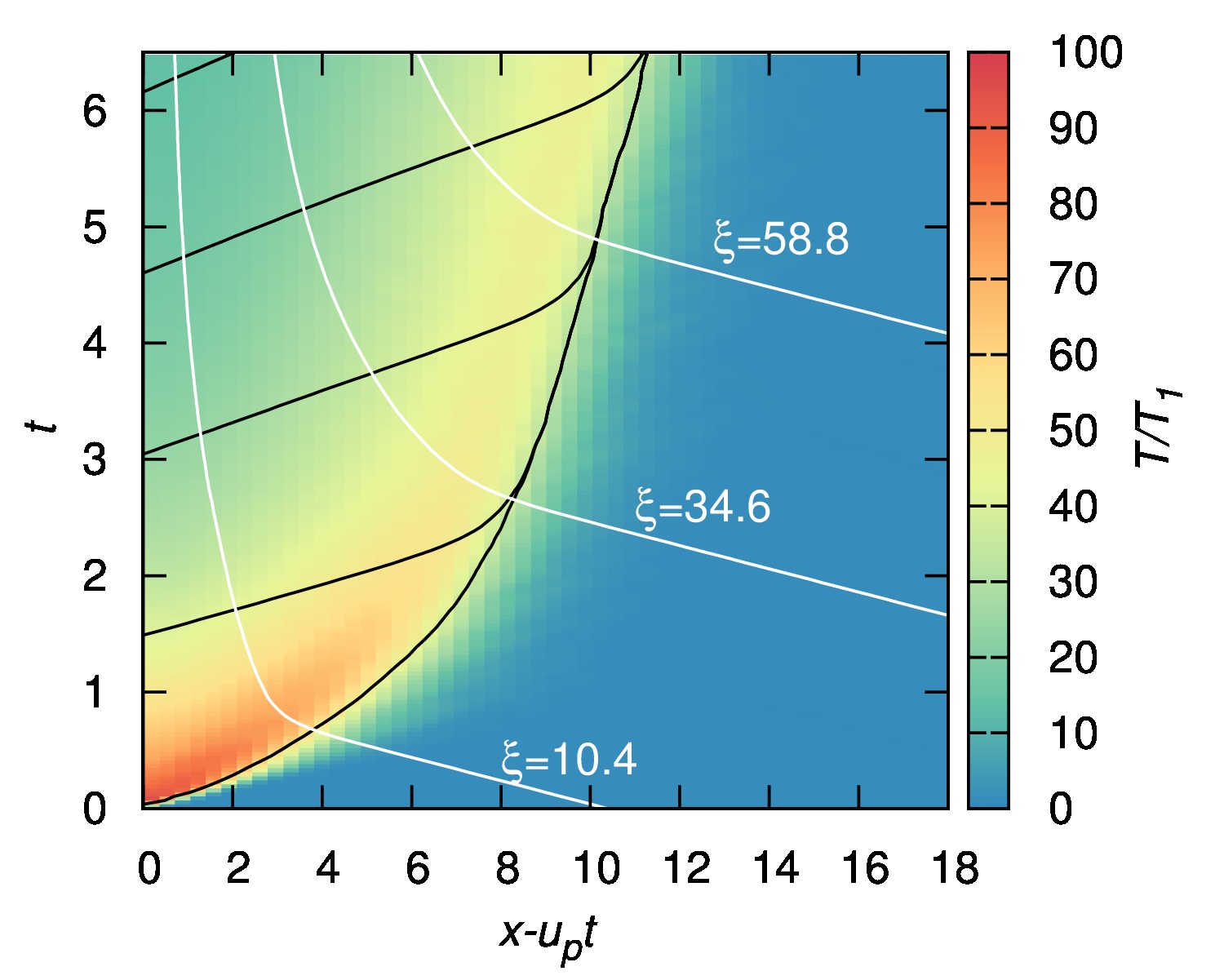}}
   	\subfloat[]{\includegraphics[trim=0.1cm 0cm 0cm 0cm, clip=true, width=0.33\textwidth]{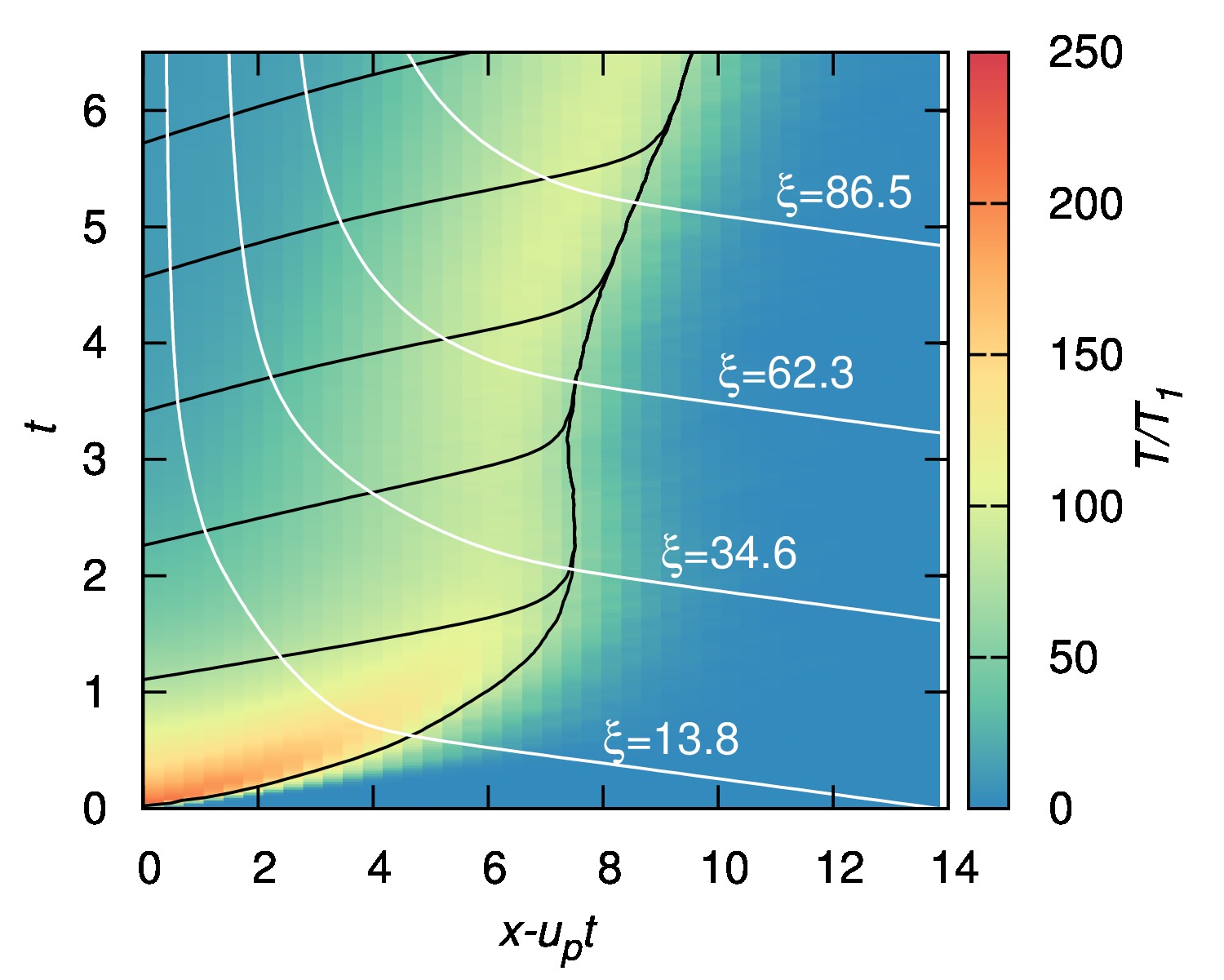}}
   	\subfloat[]{\includegraphics[trim=0.1cm 0cm 0cm 0cm, clip=true, width=0.33\textwidth]{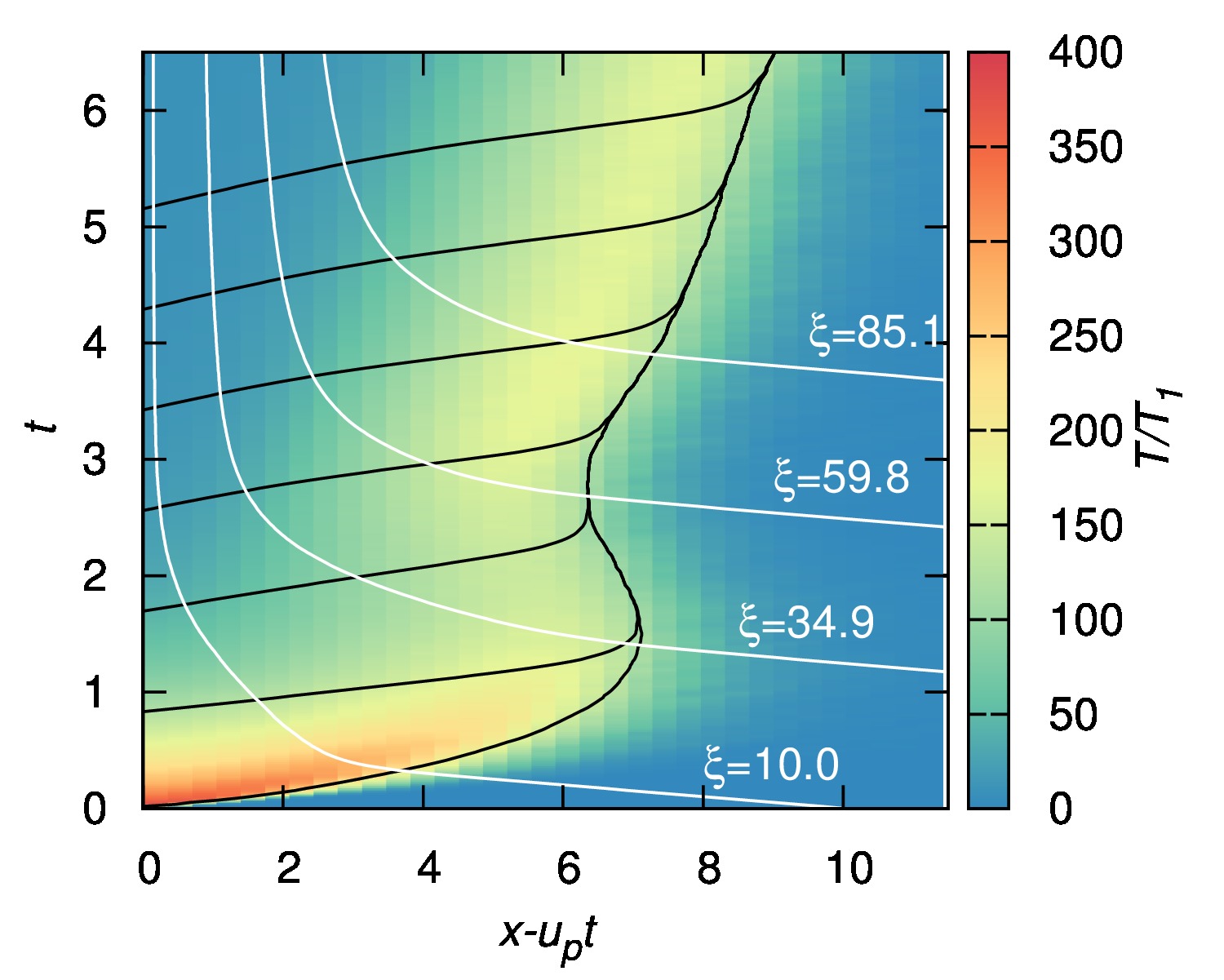}}\\\vspace{-22pt}
 	\subfloat[(a) $u_p/u^*=1.0$]{\includegraphics[trim=0.1cm 0cm 0cm 0cm, clip=true, width=0.33\textwidth]{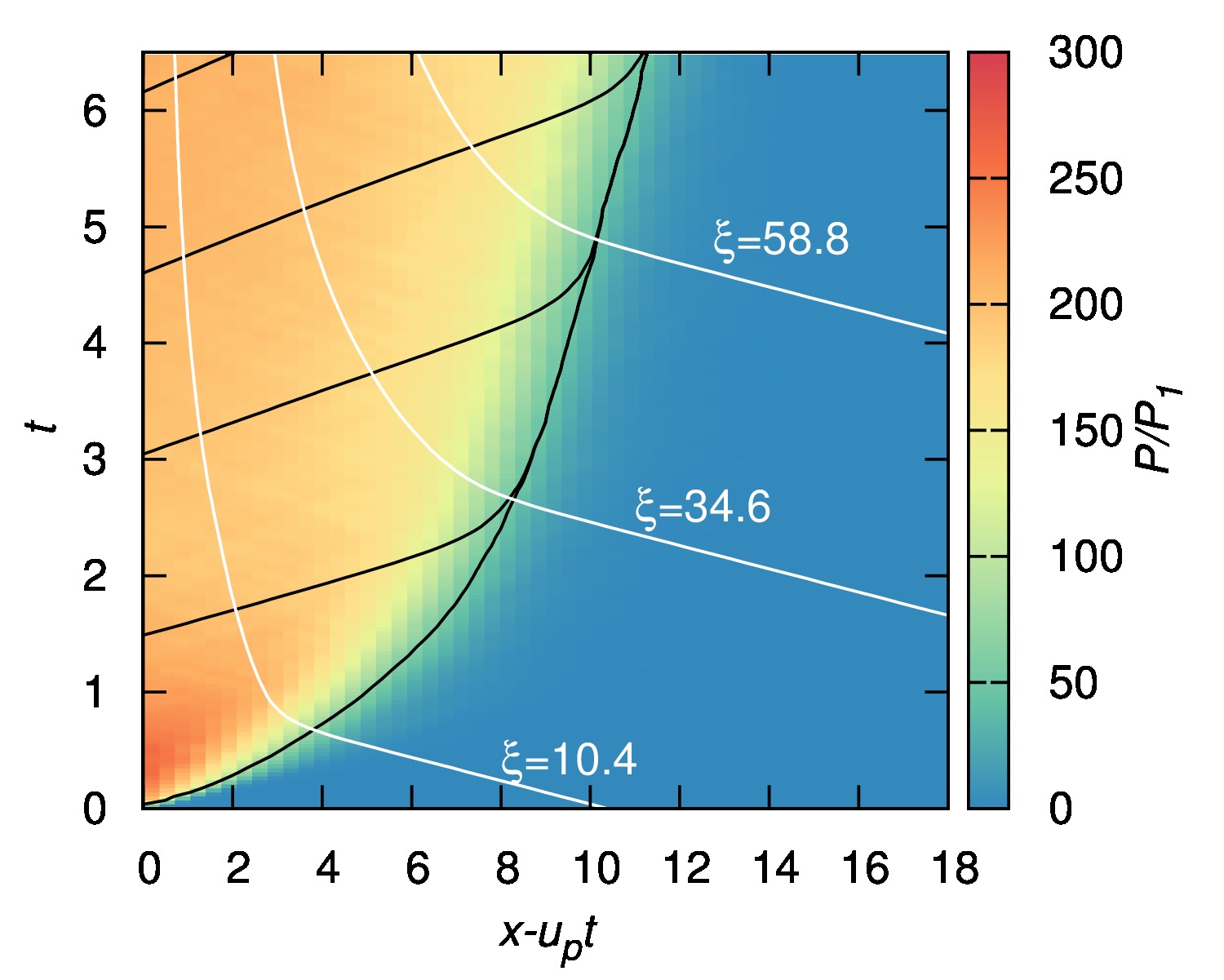}}
    	\subfloat[(b) $u_p/u^*=1.5$]{\includegraphics[trim=0.1cm 0cm 0cm 0cm, clip=true, width=0.33\textwidth]{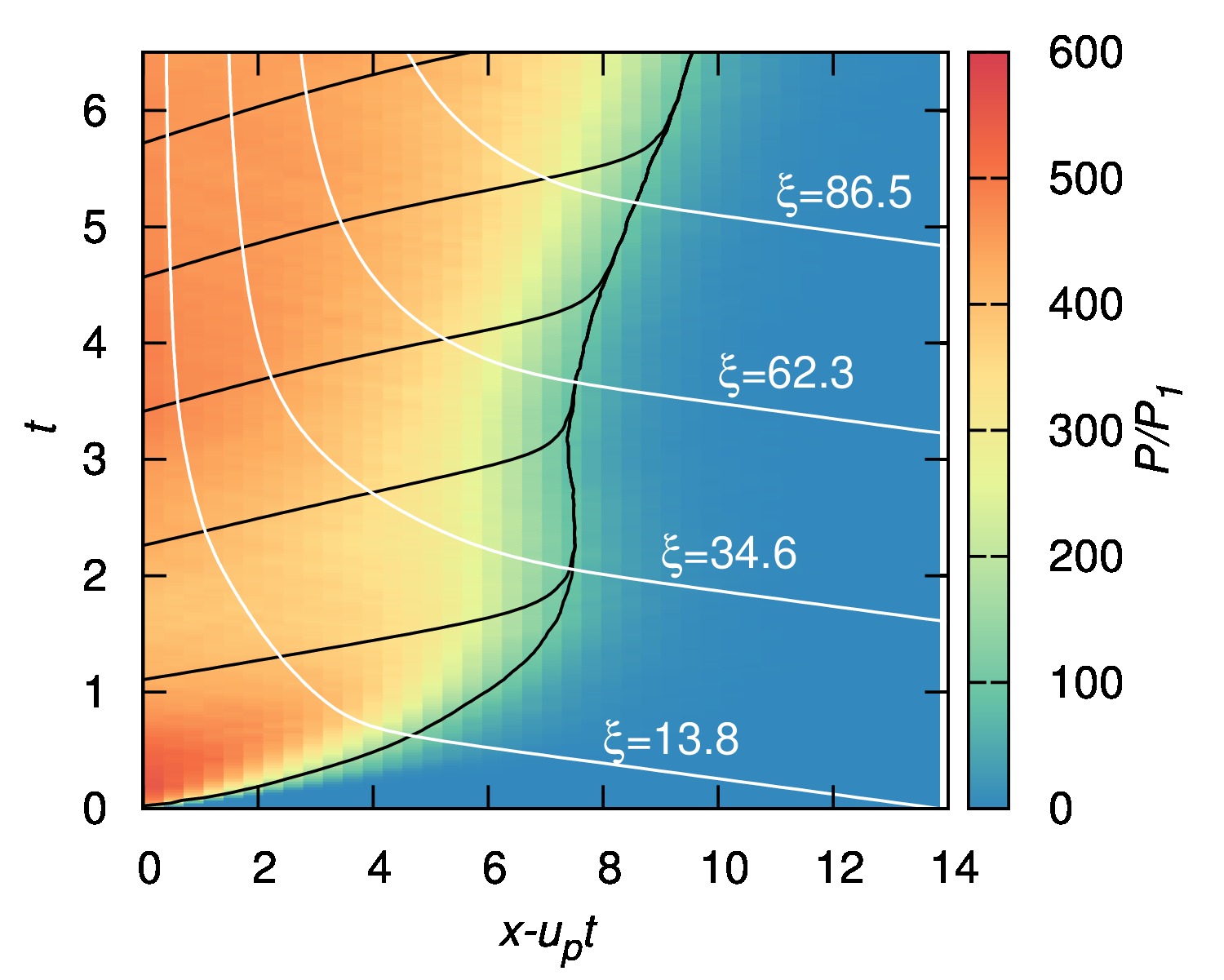}}
    	\subfloat[(c) $u_p/u^*=2.0$]{\includegraphics[trim=0.1cm 0cm 0cm 0cm, clip=true, width=0.33\textwidth]{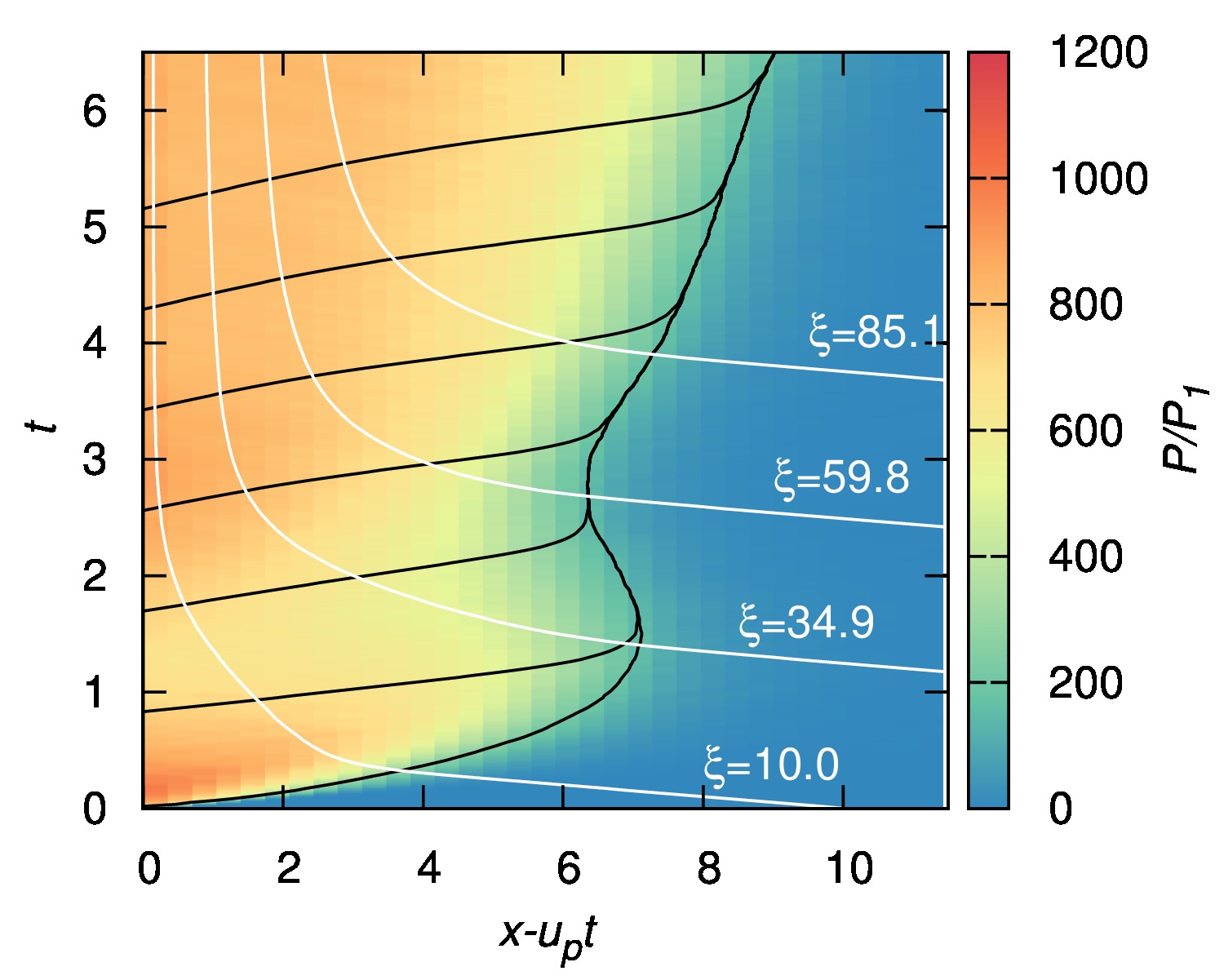}}
\caption{(Colour online) Comparison of the shock evolution for temperature (top) and pressure (bottom) obtained from MD for $u_p/u^*=$ (a) 1.0, (b) 1.5, and (c) 2.0, with $u^*=10$, $\varepsilon=0.95$ and $\eta_1=0.012$. Selected particle paths (white) and forward running characteristics (black) are shown, where $\xi=x(t=0)$.}
\label{fig:compareXT}
\end{figure*}

In order to validate the results of our dimensional analysis, we varied both $u_p$ and ${u^*}$ in the hypersonic limit $u_p \backsim u^* \gg 1$.  Figure~\ref{fig:equalratiod} and \ref{fig:equalratioT} shows results for the distributions of density and kinetic energy, respectively, after equal piston displacement while maintaining $u_p/{u^*}$=2.00 and $\varepsilon=0.95$.  Figure~\ref{fig:equalratiod} demonstrates that the distributions for density are the same after equal piston displacement. Scaling the mean kinetic energy (temperature) by the activation energy $E_A=\frac{1}{2}u^{*2}$ in Figure~\ref{fig:equalratiod}, we find the post-shock energy distributions are similar, tending towards a similar quasi-equilibrium state, where the kinetic energy tends to 5-8 \% of the activation energy. This confirms that $u_p/u^*$ is a scaling parameter for the dynamics.   In our parametric study, we henceforth maintain $u^*=10$ and vary only $u_p$ and $\varepsilon$.  We also set the initial packing factor $\eta_1=0.12$, a parameter we do not explore in the present study; see Sirmas \textit{et al}.~for its effect on the shock jump conditions in the case of non-dissipative collisions \cite{Sirmasetal2012}.  

\begin{figure*}[tbp]
\captionsetup[subfigure]{oneside,margin={0.75cm,2.5cm}} 
\subfloat[]{\label{fig:diffratiod}\includegraphics[width=.45\linewidth]{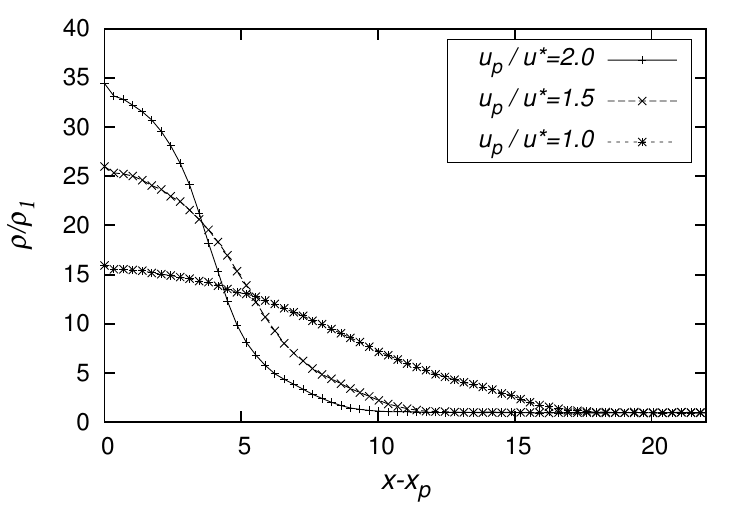}}\hfill
\subfloat[]{\label{fig:diffratioT}\includegraphics[width=.45\linewidth]{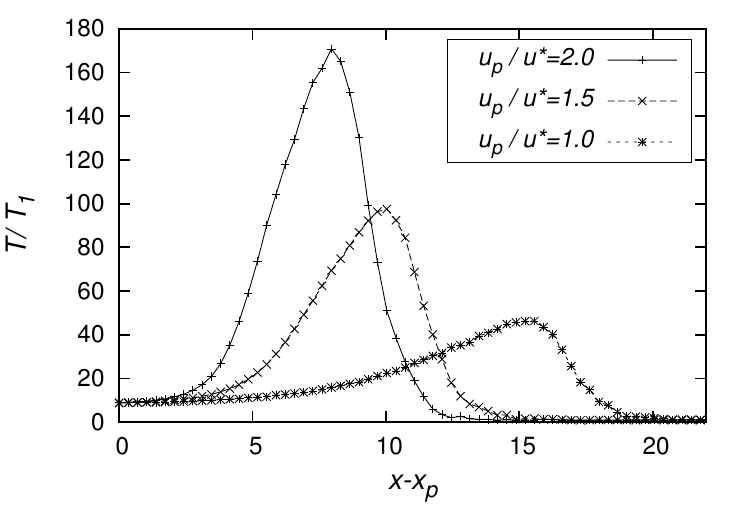}}
\caption{Ensemble and coarse grain averaged one-dimensional shock structure of (a) density and (b) temperature after a piston displacement of $x_p=138.7 \lambda_1$ for different values of $u_p/u^*$, with $\varepsilon$=0.95. }
\label{fig:diffratio_compare}
\end{figure*}
\begin{figure*}[tbp]
\captionsetup[subfigure]{oneside,margin={0.75cm,2.5cm}} 
\subfloat[]{\label{fig:ratio2diffepsd}\includegraphics[width=.45\linewidth]{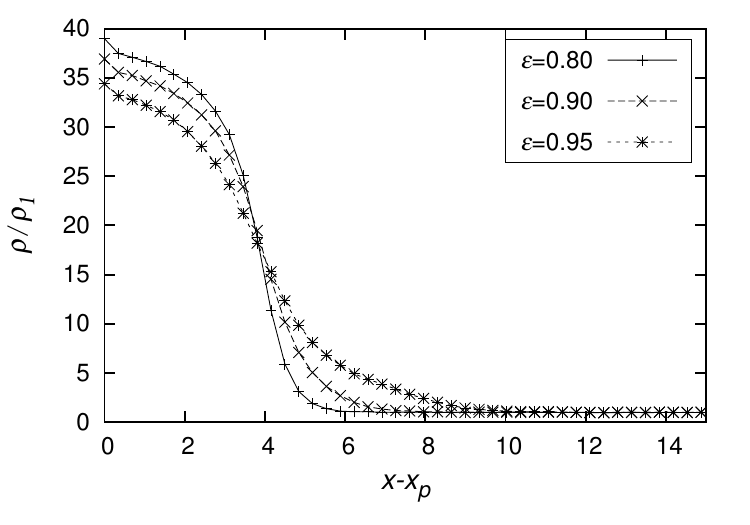}}\hfill
\subfloat[]{\label{fig:ratio2diffepsT}\includegraphics[width=.45\linewidth]{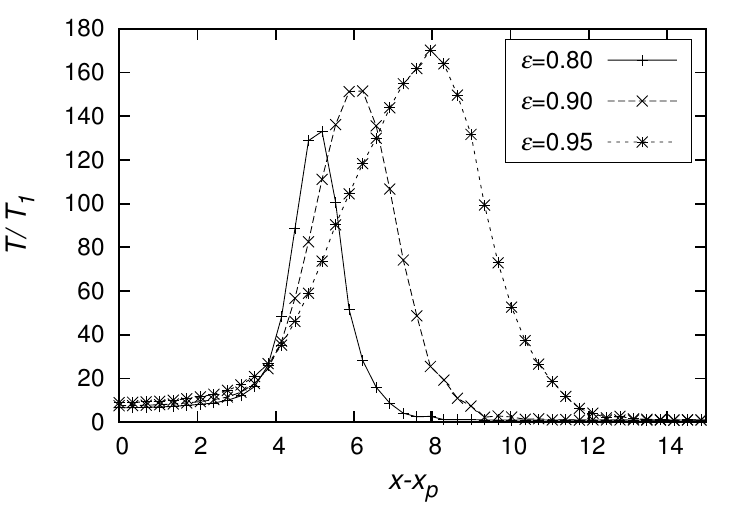}}
\caption{Ensemble and coarse grain averaged one-dimensional shock structure of (a) density and (b) temperature after a piston displacement of $x_p=156.0 \lambda_1$ for different values of $\varepsilon$ , with $u_p/{u^*}$=2.0.}
\label{fig:deffeps_compare}
\end{figure*}

\subsubsection*{Dependence on $u_p /u^*$}

It was found that $u_p /u^*$ controlled the type of dynamics observed during the relaxation process.  Figure~\ref{fig:compareXT} compares the evolution of the temperature and pressure fields obtained for  $u_p/u^*=1.0$, 1.5, and 2.0, with $\varepsilon=0.95$. For the case $u_p/u^*=1.0$ shown in Figure~\ref{fig:compareXT}(a).   In this case, the strong initial shock wave is followed by a gradual decay of the shock velocity. This decay does not cause the shock to pull back towards the piston, and the early particle paths do not experience a re-pressurization along the piston face.  When $u_p/u^*$ (Figure  \ref{fig:compareXT}(b)) is increased to 1.5, the shock front stalls with respect to the piston and a moderate re-pressurization is seen along the piston face. Further increase of the piston speed leads to a more marked shock pull-back and re-pressurization event, such as that seen in Figure~\ref{fig:compareXT}(c) for $u_p/u^*=2.0$.  The threshold for oscillatory behavior for the front shock and internal re-pressurization is approximately $u_p/u^* \approx 1$.

Figs.~\ref{fig:diffratiod} and \ref{fig:diffratioT} show a comparison of the developed distributions of density and kinetic energy after equal piston displacement for $u_p/u^*=1.0$, 1.5, and 2.0, with $\varepsilon=0.95$. Both distributions show that the distance of the shock front decreases as $u_p/{u^*}$ increases. This is attributed to the decreasing relaxation zone length for increasing $u_p/{u^*}$,  which is seen by the steeper slopes for increasing density and decreasing kinetic energy. The peak energy increases with increasing $u_p/u^*$, as expected for increasing $u_p$. All cases share a common kinetic energy at the piston face, corresponding to the quasi-equilibrium state with kinetic energy tending to 5-8\% of the activation energy.

\subsubsection*{Dependence on $\varepsilon$}
The role of $\varepsilon$ on the shock structure is to control the relaxation rate.  Figure~\ref{fig:ratio2diffepsd} and \ref{fig:ratio2diffepsT} show the distributions of density and kinetic energy, respectively, for varying $\varepsilon$ and $u_p/{u^*}=2.00$. Results show that decreasing $\varepsilon$ causes the kinetic energy to be excited and relaxed over a shorter length. This leads to a larger density gradient for lower $\varepsilon$. The peak temperature decreases as $\varepsilon$ decreases, owing to the increased dissipation during the initial excitation. The quasi-equilibrium states at the piston face show that the kinetic energies are similar, equal to approximately 5\% of the activation energy for $\varepsilon=0.80$ and 8\% for $\varepsilon=0.95$. This lower kinetic energy for decreasing $\varepsilon$ leads to a somewhat higher density at the piston face after equal piston displacement. 
\begin{figure}[htbp]
\centering
\includegraphics[width=0.9\linewidth]{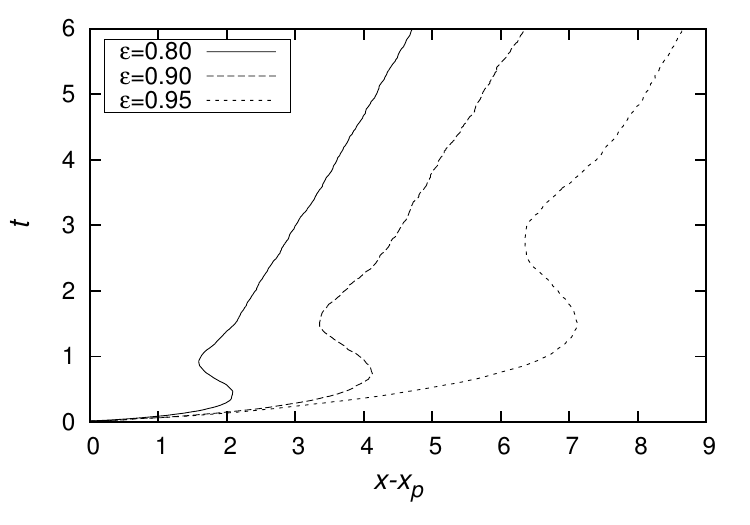}
\caption{Evolution of shock front for varying $\varepsilon$, with $u^*=10$ and $u_p=20$.}
\label{fig:diff_eps_compareMD}
\end{figure}

These trends are also seen by tracking the evolution of shock front, as shown in Figure~\ref{fig:diff_eps_compareMD}. Results show that decreasing $\varepsilon$ generates a more rapid decay of the shock front. This is shown by the shock pulling towards the piston after a shorter time. These shocks are also closer to the piston, representing a more tightly packed relaxing region. Although shocks develop faster with decreasing $\varepsilon$, all shocks tend to approximately the same developed velocity. 

\begin{figure*}[htbp]
\centering
\includegraphics[width=.8\linewidth]{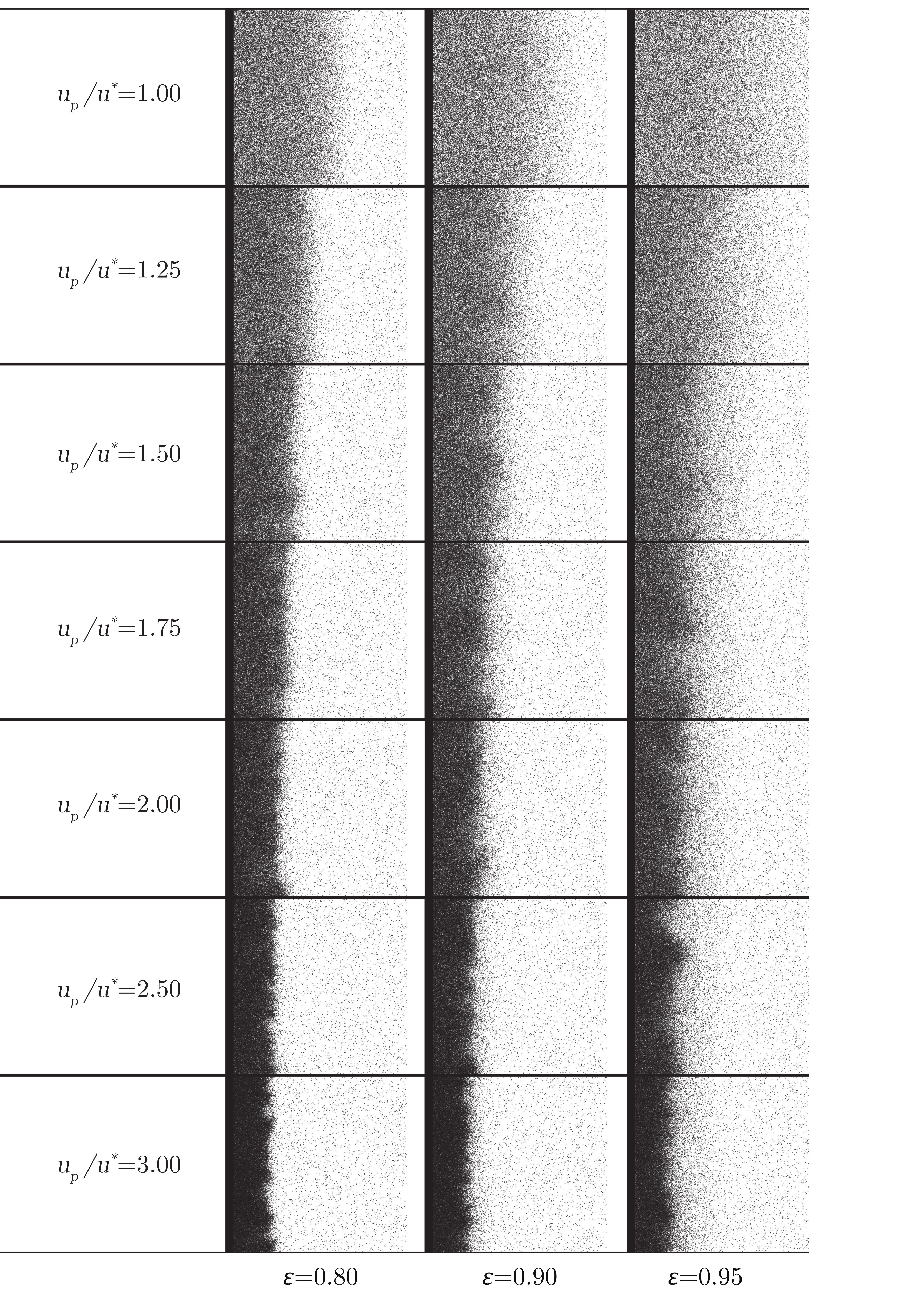}
\caption{Comparison of shock morphology for single realizations after a piston displacement of $x_p=156.0 \lambda_1$ for different values of $u_p/{u^*}$ and $\varepsilon$, where ${u^*}=10$ and $\eta_1=0.012$. }
\label{fig:appendall}
\end{figure*}

To conclude the parametric study, we look at the developed shock morphology and variation of shock instability for varying $u_p/u^*$ and $\varepsilon$. These results are shown in Figure \ref{fig:appendall} for $u_p/u^*$ ranging from 1.00 to 3.00, and $\varepsilon$ of 0.80, 0.90 and 0.95. The morphologies are taken after equal piston displacements of $x_p=156.0 \lambda_1$. Results show that the instabilities become prominent for all $\varepsilon$ with increasing $u_p/u^*$. As $u_p/u^*$ increases, the frequency of these clusters extending from the equilibrium zone increases. The number of these instabilities also increases with decreasing $\varepsilon$. We find that the wavelength of these instabilities is on the same order as the relaxation length scales, as seen in distributions presented in Figs. \ref{fig:diffratio_compare} and \ref{fig:deffeps_compare}. From these results, we see that the instabilities are noticeable for $u_p/u^*\gtrsim 1.00$, with $u_p/u^*=1.00$ difficult to discern, although this may be an artifact of the domain size.  

\subsection{C.~End States}

\begin{figure}[tbp]
 \begin{center}
\includegraphics[width=.9\linewidth]{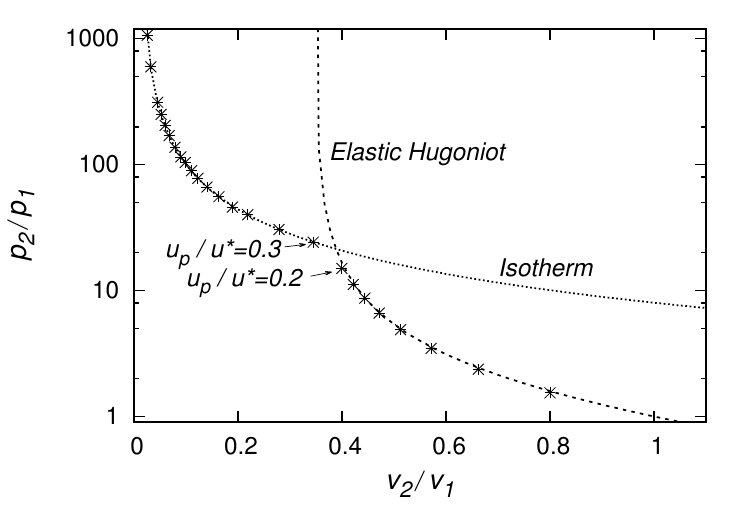}
  \caption{Results for Shock Hugoniot of equilibrium states, plotted with the elastic Hugoniot for $\eta_1=0.012$, and the isotherm corresponding to $u^*=10$.}
  \label{fig:Hugoniot}
  \end{center}
\end{figure}

The variation of the end states for different shock strengths provides the shock Hugoniot, which can be used to assess whether the shock is unstable via the BZT and/or the DK instabilities, as discussed in the introduction.  Figure~\ref{fig:Hugoniot} shows the Hugoniot curve, on a pressure-specific volume ($pv$) plane, for the case studied of ${u^*}=10$ and for $\varepsilon=0.95$.  Each point was evaluated in the post shock medium near the piston.  As discussed above, the post shock state varies very slowly after the main relaxation region (see for example Figure~\ref{fig:diffratioT}), since an exponentially small fraction of the collisions remain inelastic.  For reference, we thus register the shock state as the point where the kinetic energy is 8\% of the activation energy.  

Results show that at sufficiently small piston speeds, i.e., $u_p/{u^*}\leq 0.2$, the post-shock state follows the theoretical Hugoniot expected for a system of elastic disks, derived using Helfand's equation of state~\cite{Sirmasetal2012}:

\begin{equation}\label{eq:elasti_hugoniot}
\frac{p_2}{p_1}=\frac{\frac{1}{2}\left(1-\frac{v_2}{v_1}\right)+(1-\eta_1)^2}{\frac{v_2}{v_1}\left(1-\frac{v_1}{v_2}\eta_1\right)^2-\frac{1}{2}\left(1-\frac{v_2}{v_1}\right)}
\end{equation}
where the jump in specific volume ${v_2}/{v_1}={\rho_1}/{\rho_2}$.

 A transition occurs at approximately $u_p/{u^*}=0.2-0.3$, corresponding to a high enough piston velocity activating the inelastic collisions. Above this transition, the final state lies along the isotherm set by the activation threshold.  Using Helfand's equation of state \cite{Sirmasetal2012} for the desired isotherm, here taken as ${u_{rms}^2}/{{u^*}^2}=0.08$, the final pressure is given by:



\begin{equation}\label{eq:isotherm}
\frac{p_2}{p_1}=0.08\frac{v_1}{v_2}{u^*}^2\left(\frac{1-\eta_1}{1-\frac{v_1}{v_2}\eta_1}\right)^2
\end{equation}

\begin{figure}[tbp]
 \begin{center}
\includegraphics[width=.9\linewidth]{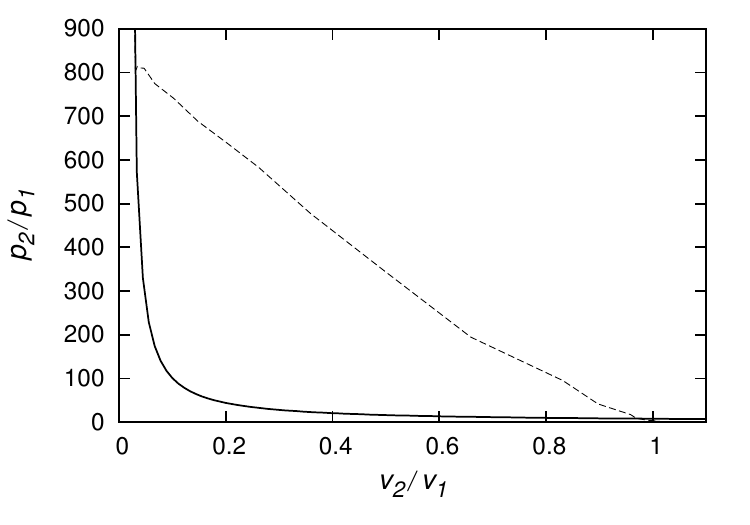}
  \caption{Results for Rayleigh line (dashed) for $u_p/u^*=2.0$ and $\varepsilon=0.95$ plotted with the Shock Hugoniot (solid).}
  \label{fig:Hugoniot_withrayleigh}
  \end{center}
\end{figure}

The evolution of the state from initial to final state across the steady shock is the so-called Rayleigh line. For further reference in our discussion of stability, Figure~\ref{fig:Hugoniot_withrayleigh} shows this path for the unstable case of $u_p/u^*=2.0$ and $\varepsilon=0.95$.

\begin{figure}[tbp]
\centering
\includegraphics[width=.9\linewidth]{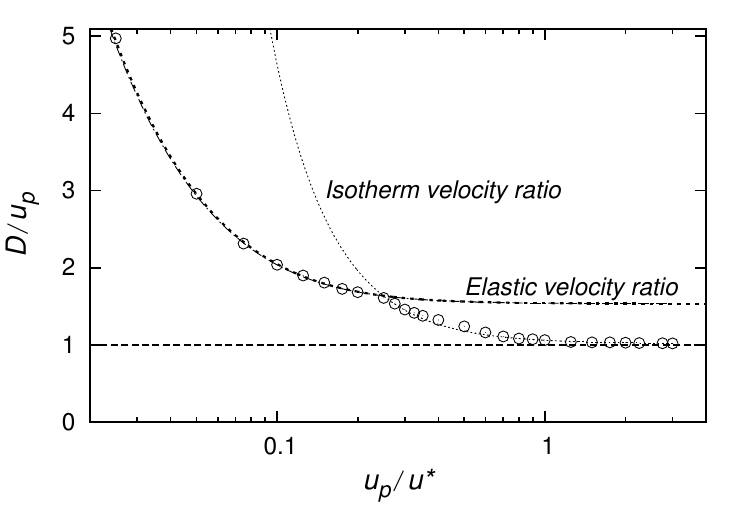}
\caption{Relationship between velocity of shock wave $D$ and piston velocity $u_p$, for ${u^*}=10$, $\varepsilon=0.90$, and $\eta_1=0.012$.}
\label{fig:SimsVelocityvsUp}
\end{figure}

The speed of the shock waves were also determined by tracking the displacement of the shock front over subsequent time intervals. Figure~\ref{fig:SimsVelocityvsUp} shows an example of the results for the shock velocity $D$ for different values of $u_p/{u^*}$ and $\varepsilon=0.90$. Results show that at the lower velocities, up to $u_p/{u^*}=0.2$, the velocities of the shock waves agree with the velocity predicted for elastic hard disks \cite{Sirmasetal2012}. The shock velocity then deviates from this ideal behaviour between $u_p/{u^*}=0.3-1.0$ until the velocity approaches $D/u_p\approx 1.0$. The shock speed is in agreement with our theoretical prediction obtained by solving the jump equations for mass and momentum with the condition of isothermicity (Eq.~\eqref{eq:isotherm}). The shock velocity is well predicted by this solution for $u_p/{u^*}>0.3$. 

To explain this transition occurring at $u_p/{u^*}=0.2-0.3$ we calculate the fraction of impact energy involved in the activated collisions, assuming a Boltzmann distribution for the state immediately behind the shock front. This is completed by following the approach used in kinetic theory to treat binary collisions, where one can begin with the rate of binary collisions per unit volume, written as \cite{Vincenti&Kruger1975}:
\begin{equation}\label{distribution}
n^2 d \frac{m}{2 kT}\exp\left\{\frac{m g^2}{4kT}\right\}g^2 \cos \psi dg d\psi
\end{equation}
This term gives the rate of binary collisions of a system of disks of mass $m$ with a number density $n$ that have a relative speed in the range of $g$ to $g+dg$, and an angle between the relative velocity and the line of action in the range of $\psi$ to $\psi+d\psi$. The impact velocity, as mentioned in Eq.~\eqref{eq:postcoll} as the normal component of the relative velocity, is $g^n=g\cos\psi$. 

Multiplying Eq.~\eqref{distribution} by $(g^n)^2=(g\cos\psi)^2$, and integrating over a range of $g^n$, yields the energy along the line of action for collisions with impact velocities within this range of $g^n$. Integrating $g^n$ from 0 to $\infty$ recovers the energy along the line of action for \textit{all} collisions. Integrating $g^n$ from $u^*$ to $\infty$ yields the energy seen along the line of action for impact velocities exceeding $u^*$. From these results, we can calculate the fraction of the average energy seen along the line of action for activated collisions, compared to that of all collisions. Acknowledging that $u_{rms}^2={2kT}/{m}$,  this ratio may be written as:

\begin{eqnarray}
\frac{{(g^n)^2_{g^n>u^*}}}{{(g^n)^2}_{g^n>0}}=
\exp\left\{-\frac{1}{2}\frac{{u^*}^2}{u_{rms}^2}\right\}\left(1+\frac{1}{2}\frac{{u^*}^2}{ u_{rms}^2}\right)\label{eq:fraction}
\end{eqnarray}

\begin{figure}[htbp]
\centering
\includegraphics[width=.9\linewidth]{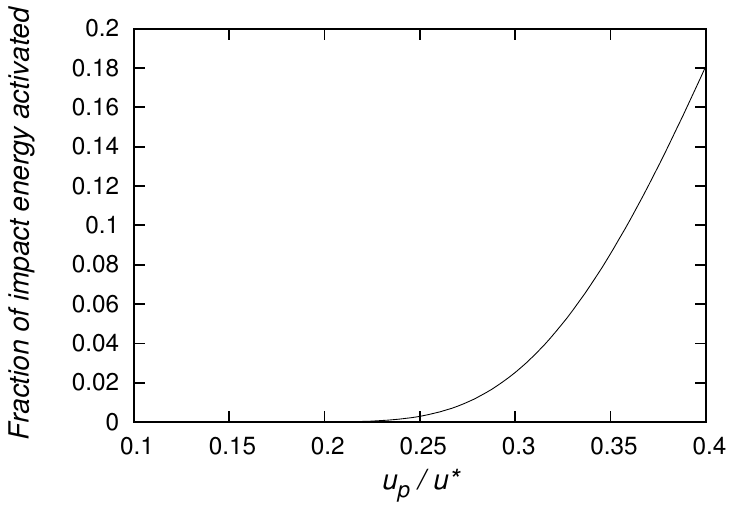}
\caption{Relationship between $u_p/u^*$ and the fraction of the impact energy involved in activated collisions behind the shock front, assuming elastic jump conditions across shock wave.}
\label{fig:fraction_activated_energy}
\end{figure}

To evaluate the difference in this ratio for $u_p/u^*=0.2$ and 0.3, we assume that the temperature at the shock jump, before noticeable dissipation, can be estimated from elastic theory~ \cite{Sirmasetal2012}, where $u^2_{rms}\approx u^2_p$. Using this equality in Eq.~\eqref{eq:fraction} allows us to approximate the fraction of impact energy that is sufficient to activate an inelastic collision. The result for this ratio near the range $u_p/u^*=0.2$ and 0.3 is shown in Figure~\ref{fig:fraction_activated_energy}. As can be seen, the fraction of impact energy that is activated is negligible for $u_p/u^*=0.2$ (0.005 \%) compared to that observed for $u_p/u^*=0.3$ (2.5\%). This clearly shows that $u_p/u^*=0.2$ is not sufficiently strong to activate a significant number of inelastic collisions, and may be approximated using elastic jump conditions. However, $u_p/u^*>0.2$ is shown to activate a more distinguishable number of collisions, which explains the transition from elastic theory seen around this value in the simulations.

\section{IV.~Discussion on Instability Mechanism}\label{sec:discussion}

\subsection{A.~Analysis of shock Hugoniot}

In the previous section, simulations showed that a shock structure does indeed become unstable with the presence of dissipative collisions. Standard explanations for shock instability are related to the shock Hugoniot \cite{Fickett&Davis1979, Landau&Lifshitz1987, Zeldovich&Raizer1966}.  For the D'Yakov Kontorovich (DK) instability, the end states lying along sections of the Hugoniot having a positive slope are expected to have a corrugation type instability~\cite{Landau&Lifshitz1987}. Figure~\ref{fig:Hugoniot} shows that the Hugoniot does not take that form, ruling out the DK instability as an influencing mechanism. 

Another possible mechanism is if the fluid is of the BZT type or undergoes phase transitions. Shock splitting is expected when the Rayleigh line, representing the state across the shock wave, intersects multiple points on the Hugoniot~\cite{Fickett&Davis1979}. Such a behavior is possible near the transition $u_p/{u^*}=0.2-0.3$ where the end state switches from lying on the elastic Hugoniot to lying on the isotherm. However, results demonstrate that it is for greater values of $u_p/{u^*}$ that the shocks become unstable. As seen in Figure~\ref{fig:Hugoniot_withrayleigh} for $u_p/u^*=2.0$, the Rayleigh line is far from this transition and does not intersect the Hugoniot in multiple locations, thus ruling out the instability associated with shock splitting. Therefore, these mechanisms can be ruled out.

\subsection{B.~Relaxation Rates and Comparison with Clustering Instability}

We now turn to another mechanism for instability previously documented for homogeneous granular gases: the clustering instability in granular gases \cite{Goldhirsch&Zanetti1993}. We wish to compare the residence time of the fluid in the shock structure and the time scale required for clusters to develop within that element of fluid.  Instability would ensue if the fluid resides within the relaxing region for longer times than required to develop the instability.  

The investigation of the clustering instability available in the literature is for a homogeneous fluid at rest, which starts cooling while kept at constant volume. The evolution of temperature before clustering is given by Haff's law (see, for example, \cite{Brilliantov&Poschel2004}). The parameters controlling this instability have been well documented \cite{Goldhirsch&Zanetti1993, Mitrano2011, Poscheletal2005}, and are not within the scope of the current work. One conclusion we will adapt is that of Mitrano \textit{et al.}: the onset of sensible clustering occurs when the evolution of granular temperature deviates by 5\% from Haff's law \cite{Mitrano2011}. Therefore by simulating the set of parameters observed in the shock waves, we can obtain the times scales for clustering necessary for comparison.

To make a comparison between the instability of the constant specific volume case and the shock case, we compare the time evolution along the particle paths traversing the shock wave structure with the time history of cooling in a constant specific volume material element.  For a meaningful comparison, this is done on time scales corresponding to the frequency of collisions, i.e., the local mean free time.  This permits to automatically avoid accounting for density changes in calculating time scales. We adopt the same criterion for onset of instability as Mitrano \textit{et al.}, and pose the question:
\textit{How many local mean free times are required for the gas to develop instability}? and \textit{How many local mean free times does the shock transition last}?  The comparison between these two time scales would permit to address whether the clustering instability plays an important role.    

To obtain the characteristic time of clustering in terms of local mean free times, we first express Haff's law in a time coordinate normalized by the local mean free time.  Haff's law expressed with time normalized by the \textit{initial} mean free time, $\tau_1$, may be expressed as~ \citep{Brilliantov&Poschel2004}:
\begin{equation}\label{eq:Haff1}
\frac{T(t)}{T_1}=\frac{1}{\left(1+t\frac{1}{4}\left(1-\varepsilon^2\right)\left(1+\frac{3}{16}a_2\right)\right)^2}
\end{equation}
where
\begin{equation}
a_2=\frac{16\left(1-\varepsilon\right)\left(1-2\varepsilon^2\right)}{57-25\varepsilon+30\varepsilon^2(1-\varepsilon)}
\end{equation}

The relation between local and initial mean free time can be shown to be 
\begin{equation}\label{eq:meantimeratio}
\frac{\tau_1}{\tau}=\frac{\lambda_1/u_{rms(1)}}{\lambda/u_{rms}}=\frac{\rho b_2(\eta)u_{rms}}{\rho_1 b_2 (\eta_1)u_{rms(1)}}
\end{equation}
Since the density $\rho$ and packing factor $\eta$ remain constant, \eqref{eq:meantimeratio} simplifies further to $\tau_1/\tau=\sqrt{{T}/{T_1}}$.

Using this change in time scales in \eqref{eq:Haff1}, we can obtain an expression for the theoretical evolution of temperature for a cooling homogeneous granular gas, in terms of time scaled by the local mean free time, i.e., $t'=\frac{t}{\tau}$. 


Constant volume clustering simulations were then conducted to determine the time when the energy of the system departs by more than 5\% from Haff's law, denoting the time for the onset of clustering $\tau_{clust}$. Since $\eta$ varies across the shock structure, packing factors ranging from $\eta=0.05-0.25$ were investigated using EDMD for $\varepsilon~=~0.8$, 0.9 and 0.95, and $N=$10,000. 

\begin{figure*}
\centering
\subfloat[$\varepsilon=0.8$]{\includegraphics[width=0.33\linewidth]{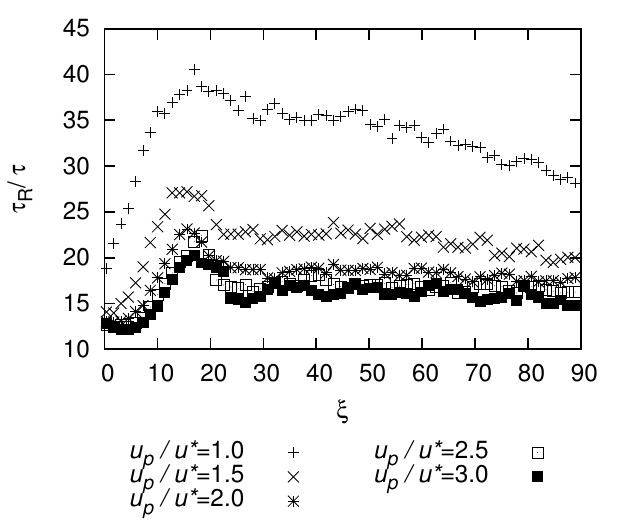}}
\subfloat[$\varepsilon=0.9$]{\includegraphics[width=0.33\linewidth]{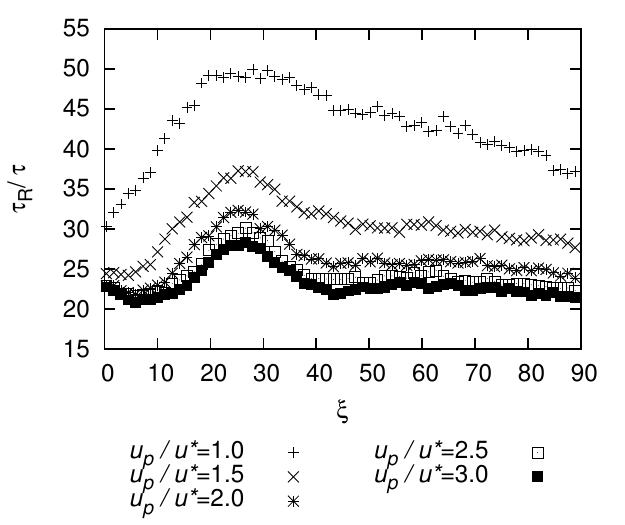}}
\subfloat[$\varepsilon=0.95$]{\includegraphics[width=0.33\linewidth]{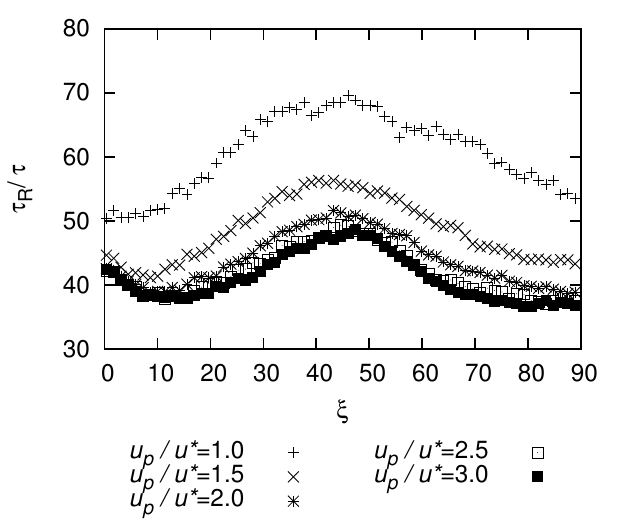}}
\caption{Exponential time constant $\tau_R$ of cooling experienced by shocked particle paths for different values of $u_p/{u^*}$ and $\varepsilon$, where $\xi=x(t=0)$.}
\label{fig:MDlocal_trelax_all_PRE}
\end{figure*}
\begin{figure*}[htbp]
\centering
\subfloat[$\varepsilon=0.8$]{\includegraphics[width=.33\linewidth]{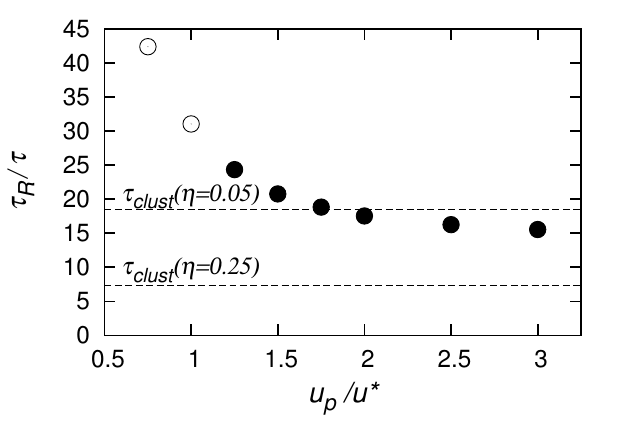}}
\subfloat[$\varepsilon=0.9$]{\includegraphics[width=.33\linewidth]{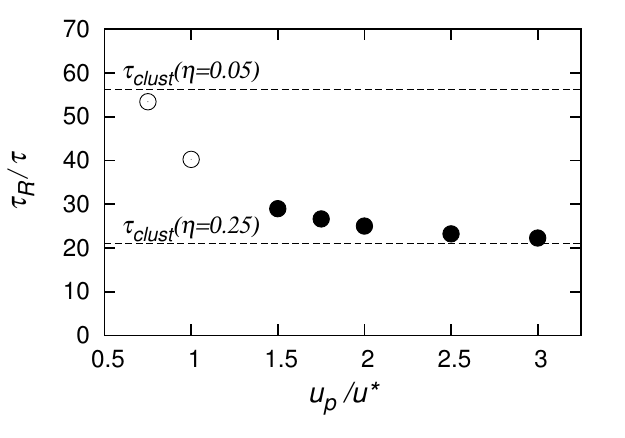}}
\subfloat[$\varepsilon=0.95$]{\includegraphics[width=.33\linewidth]{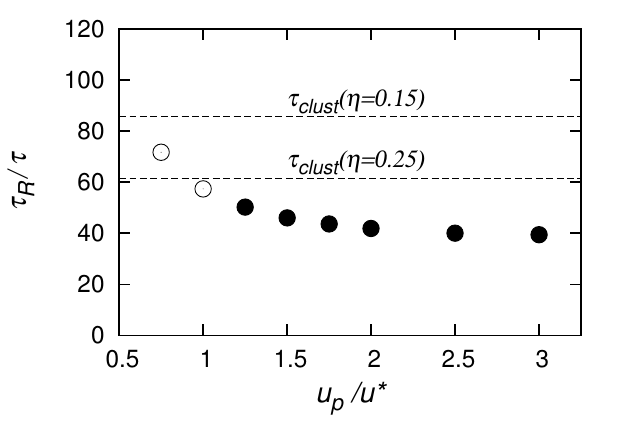}}
\caption{Mean exponential time constant $\tau_R$ of shocked particle paths for different values of $u_p/{u^*}$ and $\varepsilon$, plotted with the range in time to clustering instability for similar values of $\varepsilon$. Solid points represent simulations where unstable structures are seen.}\label{fig:timeshockrelaxcompare}
\end{figure*}

We now turn to establishing the relaxation time scale of the shocks.  We track temperature along particle paths traversing the shock structure, with time integrated by using \eqref{eq:meantimeratio} (select particle paths shown in Figure~\ref{fig:xt_pframe_MD}). The relaxation time $\tau_R$ for each fluid element is obtained by fitting the temperature decay to an exponential decay equation:
\begin{equation}\label{extc}
\frac{T(t)}{T_1}=A\exp\left(-\frac{t}{\tau_R}\right)+b
\end{equation}  

Figure~\ref{fig:MDlocal_trelax_all_PRE} shows the results of $\tau_R/\tau$ for each particle element with varying $u_p/u^*$ and $\varepsilon$. The particles generally experience fewer local mean free times to relax when $\varepsilon$ decreases or $u_p/u^*$ increases. There are variations in relaxations times seen during the evolution of the shock wave.  

Since the specific particle paths along which the instability is triggered is unknown, we compute the mean value of $\tau_R/\tau$ for each set of parameters, as shown in Figure~\ref{fig:timeshockrelaxcompare}. The results show that at higher shock strengths (higher $u_p/{u^*}$) the time constant approaches some limiting value for each $\varepsilon$. Given these results we now have a time scale to compare with the time to clustering instability $\tau_{clust}/\tau$. Since the density increases across the shock wave, the value of $\eta$ which contributes to the onset of instability can not be determined accurately. For this reason, the full range of clustering time for the range of $\eta=0.05-0.25$ is compared. 

The results shown in Figure~\ref{fig:timeshockrelaxcompare} indicate that there is practically no correlation between the observed shock instability and a residence time criterion.  Unstable shocks are generally observed when the shock relaxation time is shorter than the clustering time.  Likewise, stable shocks are observed when the shock relaxation time $\tau_R$ is longer than the characteristic time for clustering.  There is almost a perfect anti-correlation, suggesting that there is never sufficient time for a particle of fluid to develop a cluster, as it traverses the shock thickness during its relaxation process.  The results indicate that the clustering instability may not be the mechanism controlling shock instability. 

\subsection{C.~Role of initial transients on instability}

The instability of the shock was correlated above with the propensity of the relaxing medium to experience a re-pressurization event within the shock structure, as shown in Figure~\ref{fig:compareXT}.  For sufficiently small piston velocities (e.g. $u_p/u^*$=1.0) the shock wave experiences a gradual decay in strength before attaining a developed structure propagating at a constant velocity. For this shock strength we do not observe instabilities. However, as the piston velocity increases to $u_p/u^*$=1.5 and above, the shock front stalls and pulls back towards the piston for a short period before attaining a developed structure. The evolution of these stronger shock waves exhibit a re-pressurization event experienced by the early particle paths. These parameters also show the development of an unstable shock wave, suggesting a link between these initial transients and the stability. 

The re-pressurization during shock development suggests that the instability may be due to the pressure waves accelerating the flow along the piston. In this region, very strong density gradients are established.  These gradients become larger with increasing shock speed or decreasing $\varepsilon$.  These observations, and the type of instability observed with rolls forming along the density gradient, suggest that the mechanism controlling the instability is similar to Richtmyer-Meshkov or Rayleigh-Taylor type instabilities. It can be speculated that it is these wave phenomena that trigger multi-dimensional instabilities.  This is also compatible with the absence of instability, other then the original pulsation, in 1D simulations \cite{Kamenetsky_etal2000}.   Further stability analysis of this initial transient would be required, but its unsteadiness precludes using standard tools of linear analysis, such as the multi-mode approach.  

\section{V.~Conclusion}\label{sec:conclusion}
The present study showed, for the first time, that relaxing shock waves in granular gases develop instabilities, which take the form of convective rolls.  Our investigation of the possible mechanisms controlling the instabilities of shocks driven in relaxing media permitted to rule out several mechanisms.  The reconstruction of the shock Hugoniot ruled out the D'Yakov-Kontorovich instability, as well as instability related to shock splitting. Results have shown shown that the shock waves develop the instability on similar times scales as the clustering instability seen in cooling granular gases. However, away from the stability limit, the time expected for clustering to occur is found to be always larger than the time scale for relaxation across the shock, suggesting that clustering instability is not the dominant mechanism. 

Nevertheless, the onset of instability was identified during the early stages of shock development and to correspond to the sufficient condition of an internal re-pressurization of the medium and subsequent pressure wave interaction with the density gradient.  This suggests that the instability is of the Richtmyer-Meshkov type.  Further study is required to quantify the interactions.  


\section*{Acknowledgments}
We wish to acknowledge the financial support of the National Science and Engineering Research Council (NSERC)
of Canada through a Discovery Grant to M.I.R. and an Alexander Graham Bell Canada Graduate Scholarship to N.S.
\bibliography{references}

\end{document}